\documentclass[lettersize,journal]{IEEEtran}
\usepackage{amsmath,amsfonts}
\usepackage{algorithmic}
\usepackage{algorithm}
\usepackage{array}
\usepackage[caption=false,font=normalsize,labelfont=sf,textfont=sf]{subfig}
\usepackage{textcomp}
\usepackage{stfloats}
\usepackage[normalem]{ulem}
\usepackage{url}
\usepackage{tikz}
\usetikzlibrary {shapes.misc}
\usepackage{verbatim}
\usepackage{graphicx}
\usepackage{cite}
\usepackage{soul}
\usepackage{hyperref}
\usepackage{doi}
\newlength\dunder
\newcommand{\twound}{\rule{2\dunder}{0.4pt}}
\usepackage{fancyhdr}
\usepackage{kantlipsum}

\fancypagestyle{plain}{
  \fancyhf{}
  \fancyhead[C]{Conference on \LaTeX}     
  \fancyfoot[L]{This is a notice}

}
\usepackage{eso-pic}
\begin{document}
\AddToShipoutPictureBG*{%
  \AtPageUpperLeft{%
    \setlength\unitlength{1in}%
    \hspace*{\dimexpr0.5\paperwidth\relax}
    \makebox(0,-0.25)[c]{\footnotesize This article has been accepted for publication in IEEE Transactions on Signal Processing.}%
\makebox(0,-.5)[c]{\footnotesize This is the author's version which has not been fully edited and content may change prior to final publication.}
\makebox(0,-.75)[c]{\footnotesize Citation information: DOI 10.1109/TSP.2025.3574958.}
}}

\AddToShipoutPictureBG*{%
  \AtPageLowerLeft{%
    \setlength\unitlength{1in}%
    \hspace*{\dimexpr0.5\paperwidth\relax}
    \makebox(0,.75)[c]{\footnotesize\copyright~2025 IEEE. All rights reserved, including rights for text and data mining and training of artificial intelligence and similar technologies.}%
    \makebox(0,0.5)[c]{\footnotesize Personal use is permitted, but republication/redistribution requires IEEE permission.}%
    \makebox(0,0.25)[c]{\footnotesize See https://www.ieee.org/publications/rights/index.html for more information.}%
}}

\title{Fast real-time arbitrary waveform generation using graphic processing units}

\author{Juntian Tu, Sarthak Subhankar\thanks{This work was supported by the Office of Naval Research (Grant No. N000142212085), and National Science Foundation (QLCI grant OMA-2120757).
}
\thanks{Juntian Tu and Sarthak Subhankar are with Joint Quantum Institute, National Institute of Standards and Technology
and the University of Maryland, College Park, Maryland 20742 USA (email: sarthaks@terpmail.umd.edu).}
\thanks{The program code is available at https://github.com/JQIamo/AWG-on-GPU.git.}
\thanks{This work is to be submitted to the IEEE for possible publication.
Copyright may be transferred without notice, after which this version may
no longer be accessible.}
}

\maketitle

\begin{abstract}
Real-time arbitrary waveform generation (AWG) is essential in various engineering and research applications. This paper introduces a novel AWG architecture using an NVIDIA graphics processing unit (GPU) and a commercially available high-speed digital-to-analog converter (DAC) card, both running on a desktop personal computer (PC). The GPU accelerates the ``embarrassingly'' data-parallel additive synthesis framework for AWG, and the DAC reconstructs the generated waveform in the analog domain at high speed. The AWG software is developed using the developer-friendly compute unified device architecture (CUDA) runtime application programming interface  (API) from NVIDIA. With this architecture, we achieve a 586-fold increase in the speed of computing periodic radio-frequency (rf) arbitrary waveforms compared to a central processing unit (CPU). We also demonstrate two different pathways for dynamically controlling multi-tone rf waveforms, which we characterize by chirping individual single-frequency tones in the multi-tone waveforms. One pathway offers arbitrary simultaneous chirping of 1000 individual Nyquist-limited single-frequency tones at a sampling rate of 280 megasamples per second (MS/s) for a limited time duration of 35 ms. The other pathway offers simultaneous chirping of 340 individual Nyquist-limited single-frequency tones at 50 MS/s, or 55 individual tones at 280 MS/s for an arbitrary duration. Using the latter pathway, we demonstrate control over 5000-tone and 10,000-tone waveforms by chirping all of their constituent tones in groups of up to 100 tones. This AWG architecture is designed for creating large defect-free optical tweezer arrays of single neutral atoms or molecules for quantum simulation and quantum computation.
\end{abstract}


\section{Introduction}
\begin{figure*}[htb]
    \centering
    \includegraphics[width=0.8\linewidth]{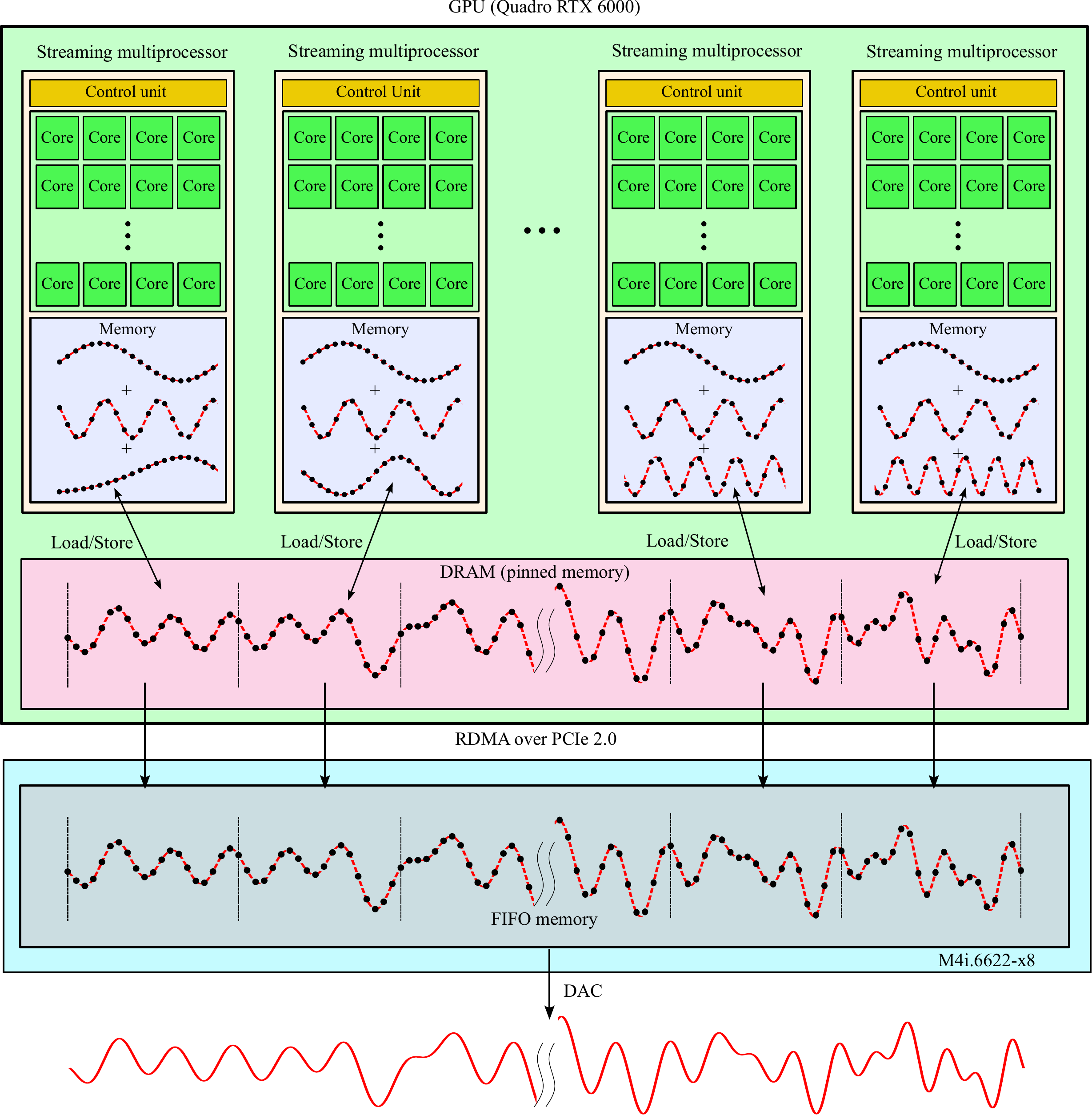}
    \caption{Illustration of the AWG architecture based on the additive synthesis framework: Each streaming multiprocessor (SM) is tasked with computing and combining chunks of the wavetables to synthesize a chunk of an arbitrary waveform that is saved in the pinned memory of the GPU. This saved arbitrary waveform chunk is transferred via remote direct memory access (RDMA) over peripheral component interconnect express (PCIe) 2.0 to the first-in, first-out (FIFO) memory and the DAC. At the DAC, this chunk is reconstructed into an analog signal.}
    \label{fig:Architecture}
\end{figure*}
\IEEEPARstart{A}{rbitrary} waveform generation (AWG) is used in a broad range of applications such as audio systems~\cite{Creasey2016}, computer music synthesis~\cite{Roads}, quantum computation and quantum simulation~\cite{Lin2019,9460084,Stefanazzi2022,Endres2016,Barredo2016,Anderegg2019,Madjarov2021,Levine2021}, electronic warfare and radio systems~\cite{NAWCWP2013}, and photolithography~\cite{Gatzen2015}. AWG is typically implemented using field programmable gate array (FPGA)-based hardware~\cite{M4i66,M4i96, TektronixAWG,KeysightAWG,ZuirchAWG,NIAWG}. FPGAs are used because of their flexibility, low-latency, high-throughput, and real-time processing capabilities~\cite{fpgaforsoftprog}. 
With the advent of accessible and general-purpose parallel programming on multithreaded multiprocessor architectures such as graphics processing units (GPUs)\footnote{Recently, general-purpose GPU (GPGPU) computing has shown promise over digital signal processors (DSPs) and FPGAs in radar signal and data processing~\cite{Perdana2017}. GPUs have shown better parallelism than FPGAs due to their large off-chip memory and high-speed coalesced access of that memory~\cite{Cong2018}. GPUs have also been used for digital signal processing \cite{Wilson2009,Pawlowski2012,Belloch2014} and software-defined radio (SDR)~\cite{Mccool2007,Plishker2011,Li2014,AKEELA2018106}. An extensive comparison between the capabilities of DSPs, FPGAs, and GPUs can be found in~\cite{HajiRassouliha2018}.}, accelerated AWG can also be performed on GPUs. In this paper, we present a GPU-accelerated AWG architecture for fast real-time control of multi-tone radio frequency (rf) waveforms essential for neutral atom quantum computation and quantum simulation.

Numerous quantum computation and quantum simulation protocols start with ordered (i.e. defect-free)  optical tweezer arrays of atoms or molecules~\cite{Holland2023,Bluvstein2024, Bao2023,Keesling2019, Eckner2023, Bornet2023}. However, the initial loading of atoms into optical tweezer arrays is stochastic, with each tweezer typically having a 50\% chance of being occupied by an atom or being empty~\cite{Fung2016,Schlosser2001,Schlosser2002,Brown2019}. An optical tweezer array is typically created by driving acousto-optic deflectors (AODs) with multi-tone rf waveforms. 
Therefore, creating a defect-free tweezer array of atoms requires rearranging the stochastically loaded array by controlling the multi-tone rf waveforms in real time~\cite{Endres2016,Anderegg2019,Madjarov2021,Levine2021}. There are two paradigms for manipulating these multi-tone rf waveforms: standard AWG~\cite{M4i66,TektronixAWG,KeysightAWG,ZuirchAWG,NIAWG} and multi-tone/parallel direct digital synthesis (DDS)~\cite{M4i96,M4i66,anUsingDDS,8103508,Young2020}.

The standard AWG paradigm involves using a central processing unit (CPU) to pre-calculate multi-tone waveforms representing static tweezer array configurations and rearrangement trajectories, saving the waveforms in memory either on the personal computer (PC) or on the FPGA-based AWG hardware~\cite{M4i66,TektronixAWG,KeysightAWG,ZuirchAWG,NIAWG}, and replaying the saved waveforms through a digital-to-analog converter (DAC)~\cite{Madjarov2021,Levine2021,Endres2016,Anderegg2019}. Although the approach is straightforward, this paradigm is intrinsically limited by the memory required to store the waveforms~\cite{KeysightAWG,anAWGModes}. 

The stochastic nature of tweezer loading requires all possible rearrangement trajectories to be pre-calculated and saved in memory. These waveforms are long because atoms cannot be accelerated arbitrarily fast without being lost from the tweezer trap~\cite{PhysRevA.109.053316}. Available laser power per tweezer limits the trap depth, and therefore limits the acceleration at which the tweezer can be moved. Fast intensity modulation (in the tens to hundreds of kHz range) during optical tweezer motion, due to the variation in the AOD diffraction efficiency over its bandwidth, can parametrically heat the atom out of the tweezer trap~\cite{Liu2019}. There are additional AOD-based cylindrical lensing effects that can weaken the trap during tweezer motion~\cite{Ricci_2024}, leading to atom loss. Therefore, rearranging stochastically loaded tweezer arrays can take up to a few milliseconds~\cite{Endres2016}. For a given laser power, a large number of tweezers implies 
lower maximum acceleration, which implies longer waveforms and therefore more memory resources required per waveform. Hence the standard AWG paradigm scales unfavorably with the number of tweezers.  
 
The multi-tone/parallel DDS paradigm, on the other hand, requires neither pre-calculation nor saving rearrangement trajectories in memory. This paradigm for generating multi-tone waveforms involves summing the outputs of individual high-bandwidth amplitude-, phase-, and frequency-tunable single-tone numerically controlled oscillators (NCOs) implemented in the fabric of an FPGA. On-the-fly changes to amplitude, phase, and frequency of the sinusoidal waveform generated by each agile NCO can be performed with a latency of less than a few microseconds~\cite{M4imanual}. For the case of one-dimensional (1D) optical tweezer arrays, each independent NCO is used to control a tweezer. Changing the NCO frequency moves the tweezer, and changing the amplitude of the NCO controls the tweezer trap depth. Therefore, dedicating each NCO per tweezer allows for full arbitrary control over 1D tweezer arrays. However, this approach also has limitations. 

The need to save each NCO wavetable (also called the phase-to-amplitude lookup table) in the FPGA fabric limits the maximum number of NCOs, with hardware resource requirements scaling unfavorably with the number of NCOs\footnote{In order to utilize the sampling rate of high-speed DACs, a few DDS cores are generally interleaved to generate one high-frequency rf tone~\cite{13357,emails,PhysRevApplied.19.054032}. Using multiple DDS cores per high-frequency rf tone increases FPGA resource usage.}~\cite{8103508,Symons_2013,emails}. Commercially available M4i.96xx and M4i.66xx cards from Spectrum Instrumentation offer a multi-tone DDS mode with a maximum of 50 and 20 tones, respectively, on only one analog output channel \cite{M4imanual}. The bespoke hardware in \cite{13357, Young2020_a} allows for up to 64 tones on one analog output channel. Therefore, scaling up to a large number of tweezers requires summing the outputs of many such modules, which can be costly. While the user-friendly software provided by Spectrum Instrumentation simplifies interfacing with their FPGA-based cards~\cite{M4imanual}, developing custom software for bespoke devices can involve significant FPGA gateware development, which can be challenging and can drive up costs~\cite{fpgaforsoftprog,HajiRassouliha2018}.

Here we develop an AWG architecture that takes inspiration from both the standard AWG and multi-tone/parallel DDS paradigms. We cast AWG as a problem highly amenable to parallelization in order to exploit multiple GPU attributes: large (tens of gigabytes (GBs)) dynamic random access memory (DRAM), high-speed access to this memory (hundreds of GB/s) by tens of streaming multiprocessors (SMs), and few thousand cores that can perform more than tens of trillions of floating-point operations per second (TFLOPs)~\cite{RTX6000}. The large DRAM on the GPU allows for the storage of thousands of wavetables with one wavetable per NCO (or tweezer). 

In the multi-tone/parallel DDS paradigm, the amplitude-, phase-, and frequency-control over the NCOs is implemented in hardware i.e. the FPGA fabric. In our AWG architecture (Fig.~\ref{fig:Architecture}), the NCO attributes are controlled in software and therefore require calculating the raw waveform data, as in the standard AWG paradigm. However, unlike the standard AWG paradigm, the waveform representing the desired NCO behavior is calculated on the fly. The waveforms from all the NCOs are summed together to generate the desired dynamic multi-tone waveform behavior. On-the-fly computation and summing NCO waveforms is reminiscent of the multi-tone/parallel DDS paradigm. 

On-the-fly computations are facilitated by the GPU's multithreaded multiprocessor architecture with high-speed access to the large DRAM and tens of TFLOPs of computation speed. The summed waveform is sent over a high-speed bus to the DAC (see Fig.~\ref{fig:Architecture}), where it is reconstructed as an analog signal that drives the AOD. Last but not least, the software was developed entirely using the developer-friendly compute unified device architecture (CUDA) runtime application programming interface (API) from NVIDIA and required no FPGA gateware development. However, our architecture has latency at the level of a few milliseconds, which is tolerable for neutral atom quantum computing and simulation due to the long coherence times of atom-based qubits~\cite{PhysRevX.14.041062,Schine2022,PhysRevX.12.021028,PhysRevX.12.021027,Picken_2019,PhysRevLett.121.123603}.

The paper is organized as follows. In Sec.~\ref{sec:theory}, we develop the theoretical framework for multi-tone AWG. In Sec.~\ref{sec:hardware}, we discuss the hardware used to implement the framework. In Sec.~\ref{sec:Operational Pathways}, we present details on the software\footnote{The code can be found at \url{https://github.com/JQIamo/AWG-on-GPU.git}.} developed for implementing the framework on hardware. In Sec.~\ref{sec:performance}, we experimentally characterize the performance of the architecture.

\section{Theory}
\label{sec:theory}

Any periodic arbitrary waveform ($V^{\text{static}}(t)$) generated from a circular buffer of length $L_s$ at a sampling clock frequency $f_s$ can be expressed as a weighted sum of the harmonics of the fundamental frequency ($f_s/L_s$)  via the Fourier transform:
\begin{align}
    V^{\text{static}}(t)=\sum_{j=1}^Na_j\sin\left(\frac{2\pi j f_s }{L_s}t+\theta_j\right),
    \label{eq:main}
\end{align}
where  $a_j$ is the amplitude  of the $j$th tone, $\theta_j$ is the phase of the $j$th tone, $f_s$ is the sampling frequency, and $N$ is the total number of tones. We ignore the DC term in the expansion as it is irrelevant in our application. Discretizing  Eq.~\ref{eq:main} in time, we get:
\begin{align}
    V^{\text{static}}[i]
    &=Q(V^{\text{static}}(t))=\sum_{j=1}^Na_j\sin\left(\frac{2\pi j}{L_s}(i \text{ mod } L_s)+\theta_j\right),\label{eq:static1}\\&=\sum_{j=1}^N a_j x^{\text{static}}_j[i \textrm{ mod }L_s]\label{eq:dataparallel}=X^{\text{static}}[i \textrm{ mod }L_s],
\end{align}
where $t=i/f_s$, $Q$ is the discretization operator, and $x^{\text{static}}_j$ is the wavetable that stores $j$ sinusoidal cycles of phase-to-amplitude mapping for the $j$th  tone. Therefore,  periodic arbitrary waveform generation can be formulated as an amplitude-weighted linear combination of $N$ wavetables ($x^{\text{static}}_j$)~\cite{Symons_2013}.
We use the phase of each tone $\theta_j$ to control for the peak/crest factor of the generated waveform $V^{\text{static}}(t)$~\cite{9301549}. The periodic multi-tone waveform $V^{\text{static}}(t)$ is used to drive an AOD to create a static optical tweezer array. It is therefore essential to suppress the crest factor using $\theta_j$ as large crest factors lead to large spikes in rf power that can damage the AOD crystal~\cite{Spence2022}.

In order to avoid undesired spurs in the spectrum of a periodic arbitrary waveform, continuous differentiability of the waveform must be enforced at the boundaries of the circular buffer. This requirement enforces a quantization on the frequency ($f_j$) of a single tone that can be represented by a wavetable buffer. A desired frequency $f$ for a tone can therefore only be approximated to its closest quantized frequency $f_j$, where     
\begin{align}
    f_j=\frac{j}{L_s}f_s=\textrm{nint}\left(\frac{f}{f_s}L_s\right)\frac{f_s}{L_s},
    \label{eq:freqapprox}
\end{align}
 and $j\in\{1,2,3\dotsc L_s/2\}\subset\mathbb{N}$ is the number of full sinusoidal cycles that can be accommodated in the wavetable buffer $x^{\text{static}}_j$, $f_s$ is the DAC sampling frequency, and nint() denotes the operation of rounding a real number to its nearest integer. For a DAC sampling rate of $f_s=280$ megasamples per second (MS/s), the error in approximating a desired frequency $f$ is less than $534$ Hz. Typical optical tweezer trap frequencies are in the hundreds of kHz range in the radial directions and tens of kHz range in the axial direction~\cite{Cooper2018,liu}. Therefore, frequency spacing between the tones in the multi-tone waveform generating the static optical tweezer array is many hundreds of kHz to a few MHz to prevent parametrically heating an atom out of a tweezer trap due to interference with light from a neighboring optical tweezer. The error in approximating the frequency in Eq.~\ref{eq:freqapprox} is therefore inconsequential in the application considered here.

The above formulation for generating arbitrary periodic waveforms (Eq.~\ref{eq:dataparallel}) can be readily extended for real-time control over the spectrum of multi-tone waveforms by separating the dynamic wavetables from the static wavetables:

\begin{align}
    V^{\textrm{dynamic}}[i]&=X^{\text{static}}[i \textrm{ mod }L_s]+ \sum_j a_j[i]\sin\left(\phi_j[i]+\theta_j\right),\\
    &=X^{\text{static}}[i \textrm{ mod }L_s]+ \sum_j a_j[i]x^{\text{dynamic}}_j[i],\label{eq:dynamicandstatic}\\
    &=X^{\text{static}}[i \textrm{ mod }L_s]+ X^{\text{dynamic}}[i],
\end{align}
where $V^{\textrm{dynamic}}[i]=Q(V^{\textrm{dynamic}}(t))$, $\theta_j$ is the phase offset propagated from the initial periodic multi-tone waveform, $x^{\text{dynamic}}_j$ stores the entire chirp trajectory, and $\phi[i]$ is the time-dependent phase during the real-time dynamic waveform generation.

We can readily program any functional form for a frequency chirp (such as linear or exponential), but here we will only consider the minimum jerk trajectory as it is typically used to minimize the heating of atoms during tweezer motion~\cite{Liu2019}. For the minimum jerk trajectory, the phase of the $j$th tone being chirped ($\phi_j[i]$) as a function of time is:
\begin{equation}
\phi_{j}[i]=
    \begin{cases}
        \frac{2\pi i}{f_s}\left(f_j^{\textrm{start}}+p[i]\right) &  0\le i\le Tf_s\\
        2\pi\frac{f_j^{\textrm{finish}}}{f_s}i+\pi T(f_j^{\textrm{start}}-f_j^{\textrm{finish}}) & Tf_s<i\le L_d\\
        2\pi\frac{f_j^{\textrm{start}}}{f_s}(i \text{ mod } L_s)+\theta'_{j} & i> L_d
    \end{cases}\label{eq:freqchirp}
\end{equation}

where  \mbox{$\theta'_j=2\pi\frac{f_j^{\textrm{finish}}}{f_s}L_d+\pi T(f_j^{\textrm{start}}-f_j^{\textrm{finish}})$} and \mbox{$p[i]=(f_j^{\textrm{finish}}-f_j^{\textrm{start}})\left(\frac{5}{2}\bigg(\frac{i}{Tf_s}\bigg)^3-3\bigg(\frac{i}{Tf_s}\bigg)^4+\bigg(\frac{i}{Tf_s}\bigg)^5\right)$}.  
Lastly, note the $i\textrm{~mod~}L_s$ indexing of only the $X^{\text{static}}$ buffer, but $i$ indexing of the ${X^{\text{dynamic}}}$ buffer. The length of ${X^{\text{dynamic}}}$ buffer ($L_d$) is
\begin{equation}
L_d=L_s\left\lceil\frac{Tf_s}{L_s}\right\rceil\ge L_s,
\label{eq:buffelength}
\end{equation}
where $T$ is the time duration for the real-time arbitrary waveform generation. $L_d$ is enforced to be an integer multiple of $L_s$ in Eq.~\ref{eq:buffelength} (see Sec.~\ref{sec:kernels} for reasons for this enforcement).

The framework presented above for controlling the spectrum of an arbitrary waveform at the single-tone level is referred to as additive synthesis~\cite{Symons_2013,Creasey2016,RUSS200999}. The multi-tone/parallel DDS paradigm is an implementation of the additive synthesis framework~\cite{Symons_2013}. In the next two sections, we will show how additive synthesis can be cast as an ``embarrassingly" data-parallel problem that can be solved efficiently on the GPU. 

\section{Hardware implementation}
\label{sec:hardware}

GPUs are a high-bandwidth, high-parallelism, high-throughput, many-core processor architecture specialized for floating-point arithmetic operations~\cite{CUDAGuide}. 
Unlike CPUs, which typically host dozens of cores and are optimized for the sequential execution of instructions via advanced pipelining and caching, GPUs host thousands of lightweight cores organized into multiple streaming multiprocessors that can execute thousands of threads concurrently~\cite{Cheng2014,Kirk2010}. Therefore, highly data-parallel problems---such as additive synthesis---can be solved efficiently on GPUs. Data-parallel programs run efficiently on vector processor-like architectures such as GPU because the same instruction is executed by all processor cores, and each processor core operates on a different data stream. Specifically, the GPU processing architecture is single instruction, multiple threads (SIMT). SIMT is different from single instruction, multiple data (SIMD) as it does not require the lockstep execution of the threads~\cite{Lindholm2008,Kirk2010,Cheng2014}.

In scientific literature, a problem is referred to as ``embarrassingly'' data-parallel when collaboration between threads is unnecessary~\cite{Kirk2010};~
 i.e., the threads work independently of each other.  
 Additive synthesis (Eqs.~\ref{eq:dataparallel}  and \ref{eq:dynamicandstatic}) is an ``embarrassingly'' data-parallel problem because it requires only two ``embarrassingly'' data-parallel tasks (see Fig.~\ref{fig:Architecture}):
\begin{itemize}
    \item \textit{Accelerated parallel computation of wavetable elements:}  The analytic nature of the phase behavior of the sine function allows for chunks of the wavetables separated in time to be computed in parallel and out-of-order by the SMs on the GPU. Additionally, the sine function is efficiently evaluated using the special function units (SFU) native to the GPU~\cite{Cheng2014}.
    \item \textit{Vector addition of amplitude-weighted wavetables:} 
    Vector addition of arrays is a classic example of an ``embarrassingly'' data-parallel problem~\cite{Kirk2010,grama2003parallel}. Eqs.~\ref{eq:dataparallel} and \ref{eq:dynamicandstatic} involve vector addition of amplitude-weighted wavetables and, therefore, can be accelerated using parallel hardware such as a GPU.
\end{itemize}

\begin{figure*}
    \centering
    {\includegraphics[width=0.85\linewidth]{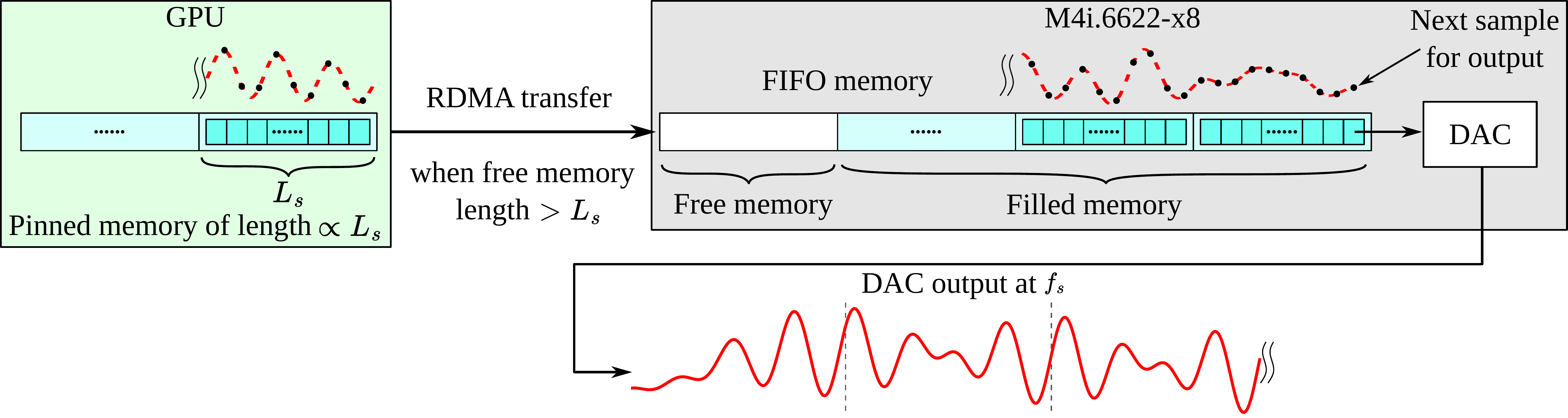}}
    \caption{Illustrating RDMA transfer between the pinned memory on the GPU and the FIFO memory on the M4i.6622-x8 card for one DAC channel. 
    }
    \label{fig:RDMA}
\end{figure*}

The additive synthesis framework is realized on an NVIDIA Quadro RTX 6000 GPU \cite{RTX6000}. Both the GPU and the M4i.6622-x8 card are connected via 
 peripheral component interconnect express
 (PCIe) slots to the desktop computer's motherboard, and are managed by the same PCIe root complex. {The desktop computer runs on an AMD Ryzen 9 5950X 16-core processor CPU, the host in this heterogeneous computing architecture. The GPU---the device in this heterogeneous computing architecture---serves as the co-processor to the CPU~\cite{Cheng2014}.} The host code on the CPU launches the GPU kernels that implement AWG (see Sec.~\ref{sec:kernels} for kernel implementations). The CPU programs the M4i.6622-x8 card and establishes the remote direct memory access (RDMA) over the PCIe 2.0 bus between the GPU's pinned memory and the first-in, first-out (FIFO) memory on the M4i.6622-x8 card (see Fig.~\ref{fig:Architecture}). The CPU also performs other tasks like running control software used in the experiment, and communicating with peripherals such as a camera for detecting images of atom arrays. 

The generated multi-tone waveform stored in the pinned GPU memory is seamlessly transferred via RDMA to the FIFO memory on the M4i.6622-x8 card in real time over PCIe 2.0 bus where it is reconstructed into an analog signal by the DAC (see Figs. \ref{fig:Architecture} and~\ref{fig:RDMA}). This RDMA data transfer pipeline is implemented using the Spectrum CUDA Access for Parallel Processing (SCAPP) software development kit from Spectrum Instrumentation\footnote{The M4i.6622-x8 card is controlled by a Virtex 6 FPGA. The role of this FPGA in the AWG architecture is to facilitate the RDMA data pipeline in Fig.~\ref{fig:RDMA}. The data pipeline is abstracted away by the SCAPP driver extension from the vendor and can be implemented using simple function calls in the CUDA API.} \cite{SCAPP,M4imanual}. The four 16-bit DACs on the M4i.6622-x8 card can run at a maximum sampling rate of 625 MS/s, when all four channels are engaged. In order to avoid buffer underrun, the FIFO buffer cannot be emptied by the DAC before it can be refilled by an RDMA transfer. To avoid buffer overrun, an RDMA transfer from the pinned memory to the FIFO memory is only invoked when the free memory length in the FIFO queue is greater than $L_s$ (see Fig.~\ref{fig:RDMA}). A DAC sampling rate less than 320 MS/s ($f_s<320$ MS/s) is necessary to avoid buffer underrun. We empirically determined the maximum DAC sampling rate by performing the following measurement. Given a filled circular pinned memory buffer, we increased $f_s$ until a buffer underrun error occurred \cite{emails}. The $f_s$ at which buffer underrun error occurred was 320 MS/s.

The optimal length of the data (or chunk) for an RDMA transfer from the pinned memory to the FIFO memory sets the lengths of all the wavetable buffers and the pinned memory buffer to integer multiples of a minimum length $L_s$. The size of this chunk is $2$ mebibyte (MiB), which implies $2^{20}$ samples per RDMA transfer (each sample is 16 bits, or two bytes, as set by the DAC resolution). When all four DAC channels are engaged, this 2 MiB RDMA chunk is multiplexed into four 0.5 MiB sub-chunks, with one sub-chunk per DAC channel per RDMA transfer. This implies an $L_s=2^{18}$ per DAC channel per RDMA transfer. For the sake of simplicity, we will assume that all four DAC output channels are engaged and all details presented in the rest of the paper will pertain to one DAC output channel.  
 
 Last but not least, as memory access latency slows down program execution, especially between the GPU memory and CPU memory in a heterogeneous computing architecture, we perform no wavetable data transfers between them. Instead, we efficiently store the computed wavetables in registers and DRAM on the GPU, facilitating rapid memory access by the GPU cores. In the next section, we elaborate on how the additive synthesis framework is implemented in software.

\section{Software Implementation\label{sec:Operational Pathways}}

\begin{figure}[h]
    \centering
    \includegraphics[width=0.85\linewidth]{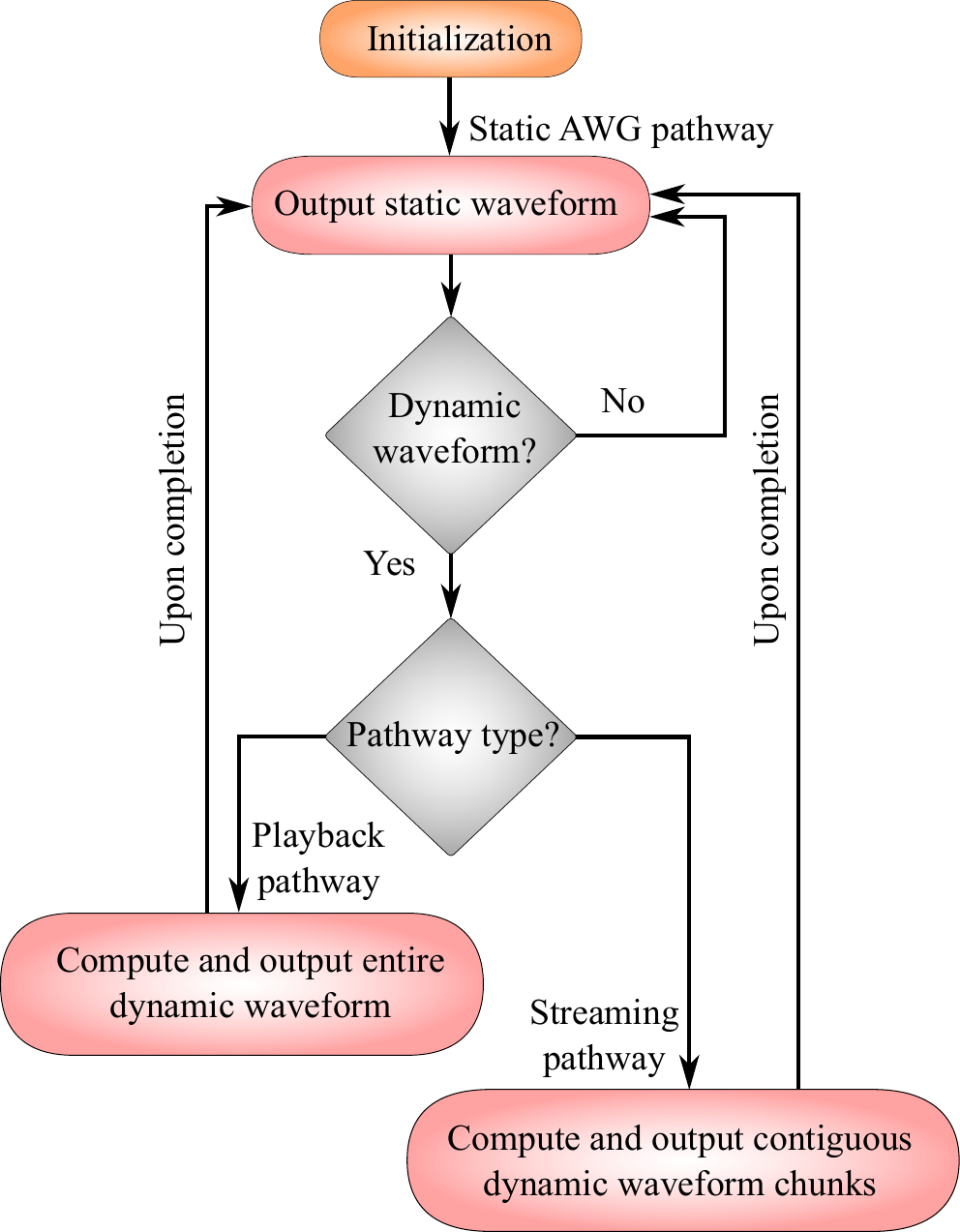}
    \caption{Three pathways are available to the user as shown in the flowchart: periodic arbitrary AWG or real-time/dynamic AWG. Both the Playback pathway and Streaming pathway are available for real-time/dynamic AWG. The Playback pathway can generate arbitrary real-time/dynamic arbitrary waveforms albeit for a short amount of time. This pathway terminates in a periodic arbitrary waveform. The Streaming pathway allows for continuous albeit limited control over the spectrum of the multi-tone waveform for an arbitrary amount of time.}
    \label{fig:High-level}
\end{figure}

NVIDIA provides the CUDA toolkit compatible with most of its GPU cards that allows parallel programming in a high-level language like C/C++. We developed the AWG software using this toolkit. In this section, we elaborate on waveform generation pathways implemented using additive synthesis. One pathway is devised for generating periodic arbitrary waveforms (see Sec.~\ref{sec:Static arbitrary waveform synthesis}), and two pathways are devised for generating dynamic/real-time arbitrary waveforms (see Sec.~\ref{sec:Dynamic arbitrary waveform synthesis}). 
A flowchart representing the control flow of the software is shown in Fig.~\ref{fig:High-level}. All pathways share a common initialization state in which the system resets the GPU, configures the M4i.6622-x8 card, and allocates the memory for the wavetable buffers in the GPU DRAM. 

\subsection{Static AWG pathway\label{sec:Static arbitrary waveform synthesis}}

This pathway is named as such because it is used to generate an optical tweezer array where all atoms are static. The periodic arbitrary waveform data $V^{\text{static}}[i]$ is evaluated only once and then saved to the pinned memory buffer. We generate a periodic arbitrary waveform at the DAC $V^{\text{static}}(t)$ by repeatedly transferring $V^{\text{static}}[i]$ to the FIFO memory via RDMA over PCIe 2.0.

\begin{figure*}[ht]
    \centering
    \includegraphics[width=\linewidth]{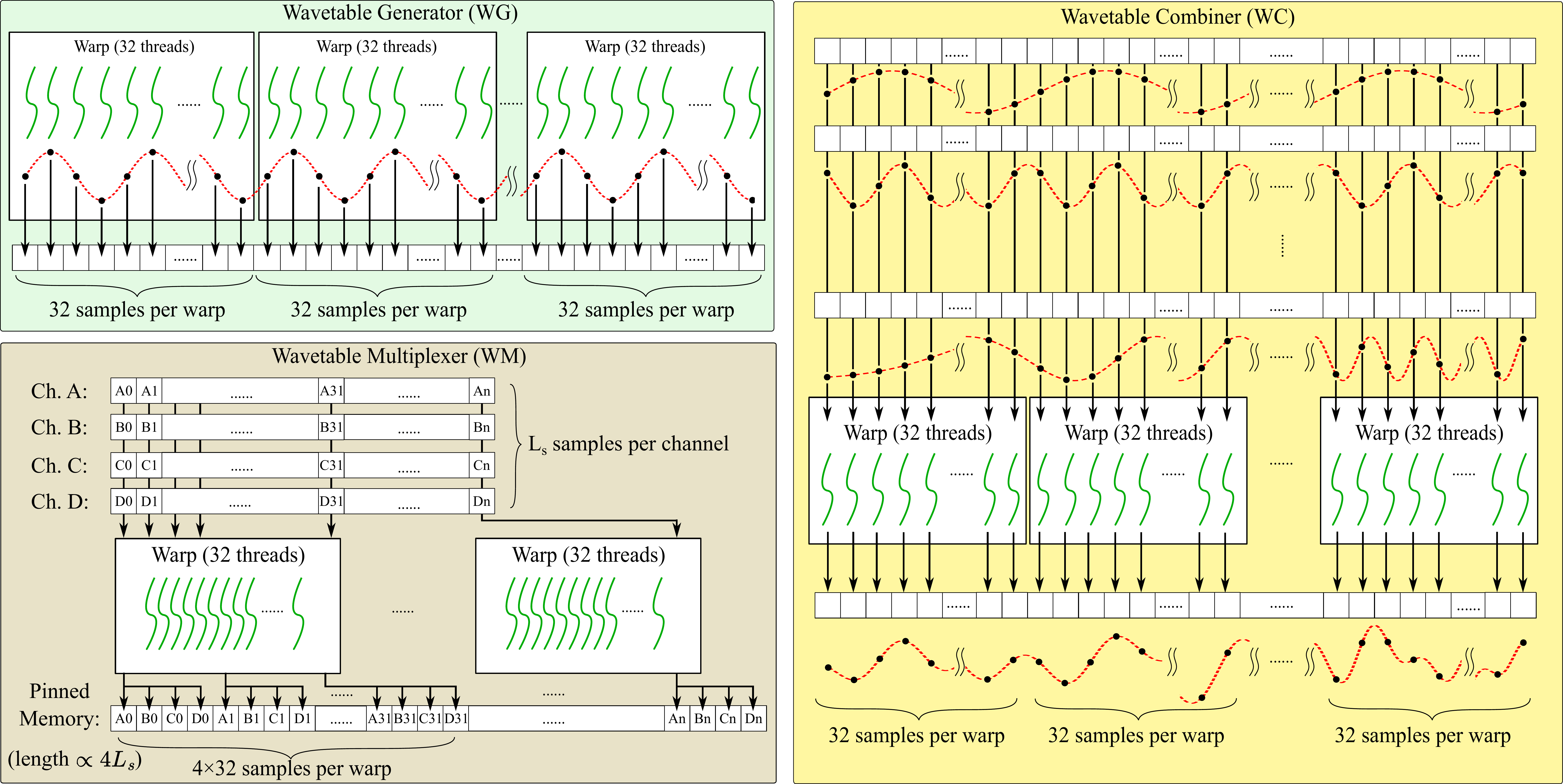}
    \caption{Illustrating the kernel implementations: In the wavetable generator kernel (WG), each thread samples and evaluates a point in the analytic mathematical expression for a tone (like a sine or a chirp) and saves it to the tone's wavetable buffer; the wavetable combiner kernel (WC) vector-adds the wavetables generated by the WG; the wavetable multiplexer kernel (WM) restructures the arbitrary waveform data for the four DAC channels so that each DAC channel can reconstruct its own arbitrary waveform data.}
    \label{fig:Kernels}
\end{figure*}

\subsection{Dynamic AWG pathways\label{sec:Dynamic arbitrary waveform synthesis}}

We implement two pathways for dynamic waveform synthesis. In what we call the ``Playback'' pathway, the entire real-time waveform data $V^{\text{dynamic}}[i]$ is computed and saved to the pinned memory buffer prior to transferring it to the FIFO memory via RDMA and then to the DAC. This pathway enables complete control over the spectrum of an arbitrary waveform $V^{\text{dynamic}}(t)$ for a brief period of time, facilitating a highly complex transformation of an initial periodic multi-tone waveform to a final periodic multi-tone waveform. For instance, a static tweezer array of stochastically loaded atoms can be rearranged into a defect-free static tweezer array of atoms using this pathway. As the calculation is performed on the GPU in real-time, there is a latency between the start of the computation and when the DAC is ready to output the computed waveform. The delay increases with the complexity and duration of the dynamic waveform. 

The other pathway is called the ``Streaming'' pathway. In this pathway, each subsequent waveform chunk is evaluated in the time interval between the current and upcoming RDMA transfers. This allows for continuous streaming of the real-time arbitrary waveform $V^{\text{dynamic}}(t)$. The waveform chunk computation must be performed faster than the time it takes to perform an RDMA transfer. While one can slow down $f_s$, the Nyquist frequency is also reduced proportionally. Large Nyquist frequencies require fast computations, and this limits control over the spectrum of the arbitrary waveform to only a few tones at a time. Unlike the Playback pathway, the Streaming pathway can be used to stream waveforms indefinitely. 

\subsection{CUDA kernel implementations\label{sec:kernels}}
The waveform synthesis pathways are implemented using CUDA kernel functions, or simply kernels, which are user-defined functions executed by the SMs on GPUs. CUDA kernels are executed in 32-thread groups called warps. A warp is the unit of thread scheduling on a GPU~\cite{Kirk2010}. 
Multiple warps are grouped together into blocks, which are then assigned to the SMs for execution~\cite{Cheng2014}. All 32 threads in a warp share a common instruction register, but process on different data. This is the SIMT parallel programming architecture. Furthermore, each thread in a warp has its own set of registers in addition to its local memory and additional memory that it shares with other threads in the warp called shared memory. 
In order to achieve massive speedup through data parallelism, the data should ideally be divided into warp-sized groups of 32, where each thread works on one piece of data independently. It also helps greatly if there are no control flow divergences in the warp~\cite{Kirk2010}. In our case, the length of all buffers is an integer multiple of $L_s=2^{18}$, which is a multiple of 32. 

All kernels are designed to launch $L_s$ threads. The threads are indexed using the variable $k$, where $k\in[0,L_s-1]$. 
The length of each static wavetable $x^{\textrm{static}}_j$ is equal to the total number of threads. The length of each dynamic wavetable $x^{\textrm{dynamic}}_j$ is $L_d$, which is an integer multiple ($L_d/L_s$) of the total number of threads (see Eq.~\ref{eq:buffelength}). The number of threads is orders of magnitude greater than the number of available cores, which helps increase GPU throughput via latency hiding by assigning multiple thread blocks per SM.

There are no control flow divergences in our kernel implementations for the static AWG pathway. Control flow divergence can occur in the dynamic pathway. When the frequencies of the tones are chirped via the minimum jerk trajectory (Eq.~\ref{eq:freqchirp}), only 1 warp out of the $L_d/32$ warps can exhibit control flow divergence. This warp processes data with array index $k\in[Tf_s-x+1,Tf_s+32-x]$, where $x$ threads process data with index $k\in[Tf_s-x+1, Tf_s]$ (in parallel) followed by $32-x$ threads processing data with index $k\in[Tf_s+1, Tf_s+32-x]$ (in parallel). {This control flow divergence will impact $(1~\textrm{warp})/(L_d/32~\textrm{warps}) \le 1/(L_s/32)$ of the total execution time, which evaluates to no more than $2^{-13}$ for $L_s = 2^{18}$ (see Eq.~\ref{eq:buffelength})~\cite{Kirk2010}. Therefore, the performance drop from the control flow divergence is minimal. The control flow divergence occurs when $Tf_s\not\equiv0 ~(\textrm{mod }32)$, but disappears when $Tf_s\equiv0 ~(\textrm{mod }32)$.} Control flow divergence cannot occur if all threads in a warp take the same control path~\cite{Kirk2010}. We have implemented several other strategies to speed up the code performance, which can be found in the appendix. 

There are three types of CUDA kernels implemented in the program: wavetable generator (WG), wavetable combiner (WC), and wavetable multiplexer (WM) as illustrated in Fig. \ref{fig:Kernels}. The WG kernel computes the wavetables. For computing single-frequency tone wavetables, the kernel allocates the $k$th thread to compute $x^{\textrm{static}}_j$ wavetable data at index $k$. This thread index-to-wavetable index mapping facilitates high data parallelism as the $k$th thread independently computes a single data point in the wavetable buffer~(see Fig.~\ref{fig:Kernels}). For computing dynamic wavetables, the $k$th thread computes $L_d/L_s$ data points for each $x^{\textrm{dynamic}}_j$ wavetable at array index $k+mL_s$, where $m\in[0,L_d/L_s-1]$. 

The WC kernel performs the vector additions (or subtractions if necessary) of the saved wavetables. The thread indices are mapped to wavetable indices in the same way as described in the WG kernel paragraph above. When the kernel is launched, the $k$th thread initializes one of its registers with zero in floating-point format, and then iteratively adds or subtracts the values in the wavetables at index $k$ to this register~(see Fig.~\ref{fig:Kernels}). This implementation accesses each data point in the wavetables only once, thus mitigating the latency associated with frequent global memory access.

The WM kernel multiplexes the arbitrary waveform wavetables for the four DAC channels into a single buffer, which is transferred to the FIFO memory through RDMA in $4L_s-$sized chunks (see Fig.~\ref{fig:Kernels}). The data in the FIFO memory is read sequentially in groups of four samples in a single clock cycle. Each group contains the wavetable values for four DAC channels. The thread index to wavetable index mapping is also natural for the WM kernel. The $k$th thread takes the $k$th elements from each of the four wavetable buffers and arranges them sequentially as a group of four in the pinned memory buffer. Due to the simplicity of the kernel, it takes, on average, tens of microseconds to execute and doesn't affect the program performance. Note that the computations implemented by the WG, WC, and WM kernels are ``embarrassingly'' data-parallel as each thread executes independently of any other~(Fig. \ref{fig:Kernels}).

In practice, these kernels are modified to match the use case and merged together to minimize high-latency operations, such as repeatedly accessing the same chunk of data from DRAM memory. In the static AWG pathway, the kernels are launched sequentially. The {WG and WC kernels are merged together to generate and store in memory} all static (individual and summed) wavetables that will be reused in the dynamic pathways.
Lastly, the WM reorganizes the data and saves it in the pinned memory buffer before it is transferred via RDMA to the DAC output channels for analog reconstruction.

In the Playback pathway, the WG and WC are {also }merged into a single kernel in order to compute the entire dynamic waveform. Upon completion of the WG+WC kernel, the WM kernel reorders the dynamic waveform data before sending it over to the DAC output channels. This pathway ends in a periodic arbitrary waveform.

In the Streaming pathway, the WG, WC, and WM are merged into a single kernel as new arbitrary waveform data is generated on the fly and streamed to the DAC output channels via RDMA. 

Lastly, shared memory usage can boost program performance. Shared memory is a programmer-controlled memory resource accessible to all threads in a block. Shared memory has much lower capacity compared to global memory, but provides significantly faster access bandwidth~\cite{CUDAGuide,Cheng2014,Kirk2010}. In our implementation, the shared memory stores the waveform parameters such as the frequencies, amplitudes, and phases of all chirping tones. In the WG kernels, the parameters of all chirping tones need to be accessed by each thread. Instead of repeatedly accessing the data from global memory, we pretransfer the data into the shared memory for each block, where the threads in the block can access the data efficiently. As a result, the shared memory reduces the latency arising from frequent global memory access, especially when a large number of tones need to be chirped. The efficacy of shared memory is limited for the WC and WM kernels due to the fact that the data is accessed in a coalesced manner from global memory, data is not reused by other threads during kernel execution, and each thread operates on a single wavetable index at a time.

\section{ Performance\label{sec:performance}}

 In this section, we evaluate the performance of the static and dynamic pathways implemented using additive synthesis. We also present details on the computational latency, memory usage, and power consumption.
\subsection{Static AWG pathway}

\begin{figure*}
    \centering
    \subfloat[\label{fig:chirp2a}]{\includegraphics[width=0.31\linewidth]{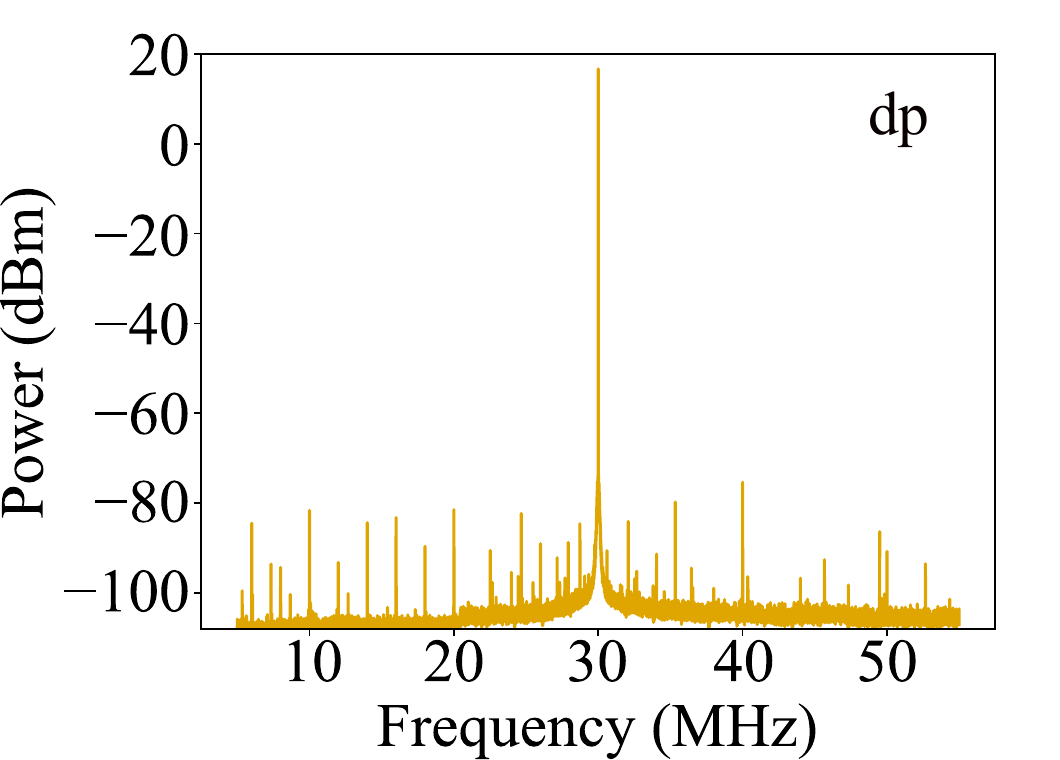}}
    \qquad
    \subfloat[\label{fig:chirp2b}]{\includegraphics[width=0.62\linewidth]{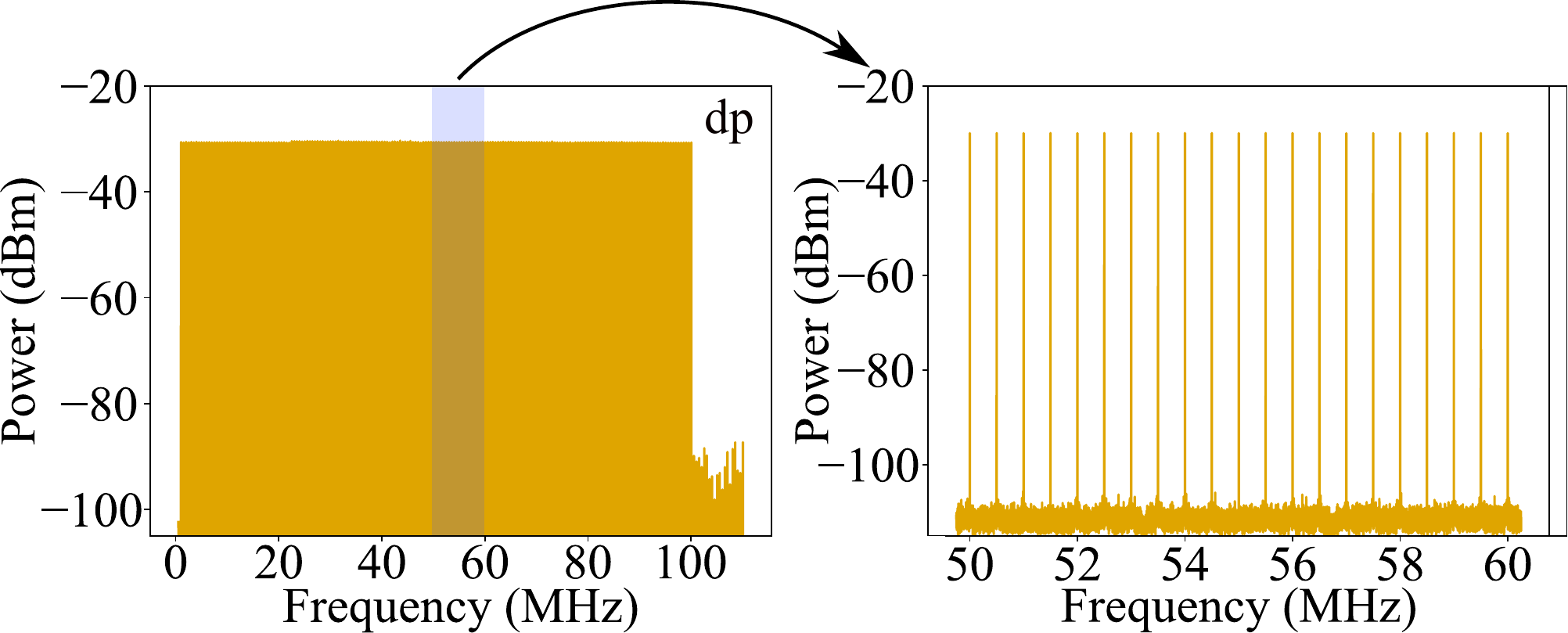}}
    \hfill
    \subfloat[\label{fig:chirp2c}]{\includegraphics[width=0.31\linewidth]{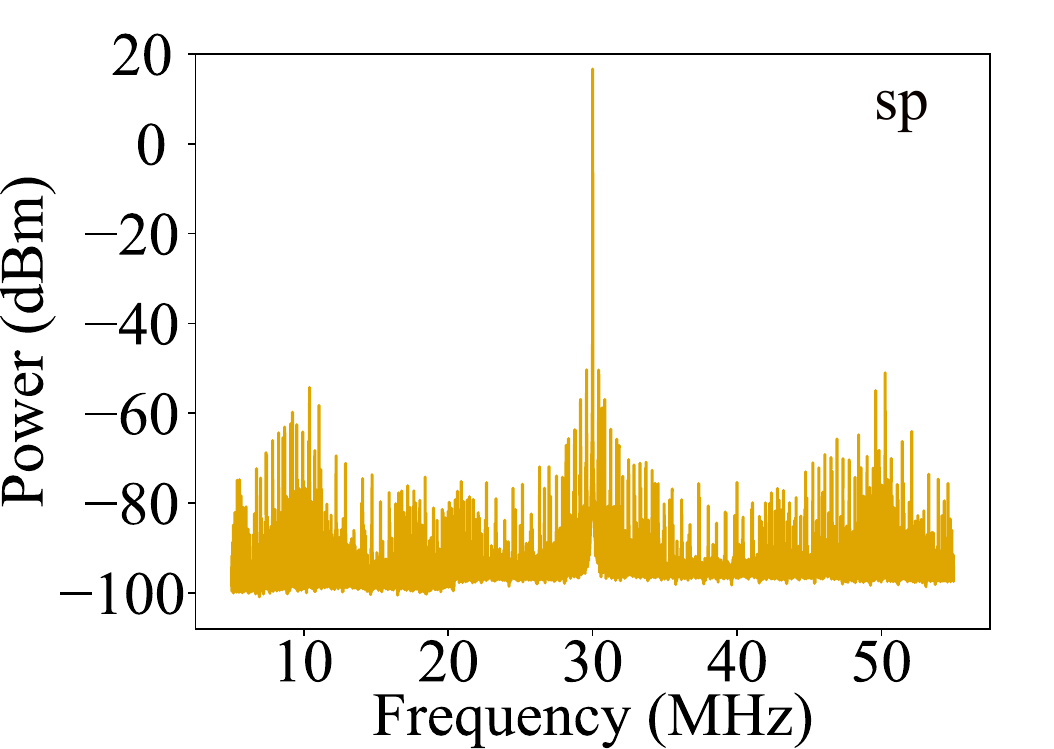}}
    \qquad
    \subfloat[\label{fig:chirp2d}]{\includegraphics[width=0.62\linewidth]{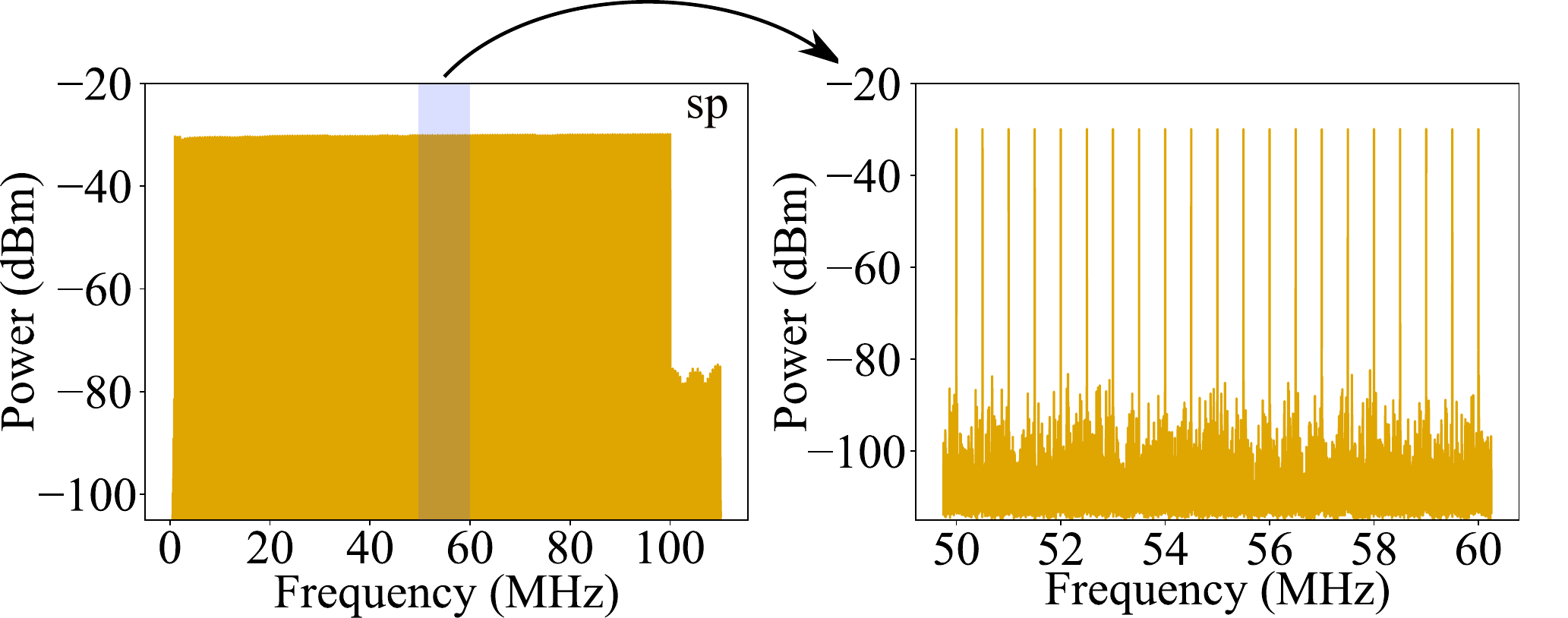}}
    \caption{The spectrum of: (a) a single-tone sinusoidal waveform with wavetable data computed in dp, (b) a 199-tone ``defect-free" waveform with wavetable data computed in dp, (c) a single-tone sinusoidal waveform with wavetable data computed in sp, (d) a 199-tone ``defect-free" waveform with wavetable data computed in sp\\
   }
   \label{fig:waveforms}
\end{figure*}

\begin{figure}[h]
    \centering
    \subfloat[\label{fig:staticL}]{\includegraphics[width=\linewidth]{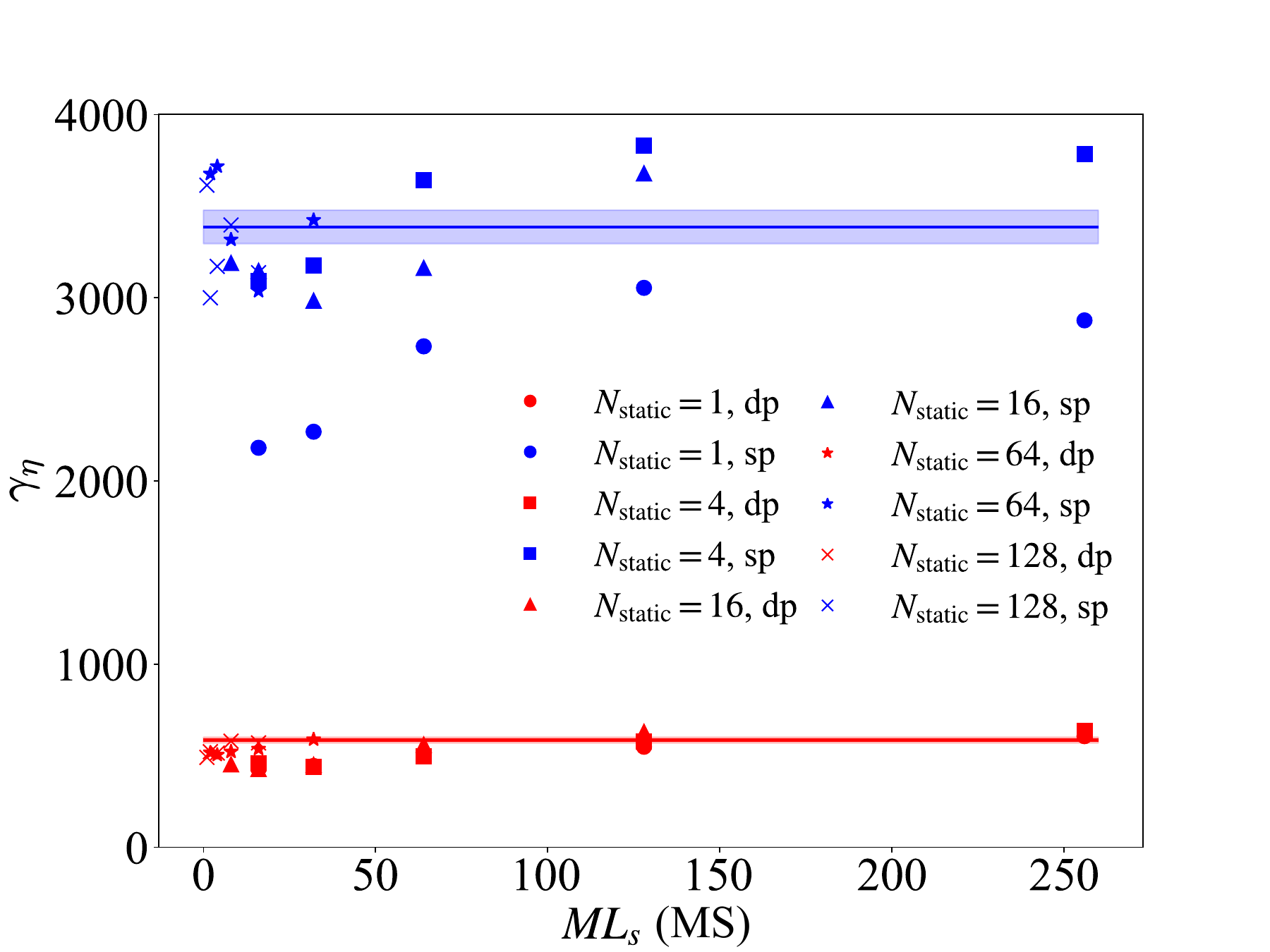}}
    \qquad
    \subfloat[\label{fig:staticN}]{\includegraphics[width=\linewidth]{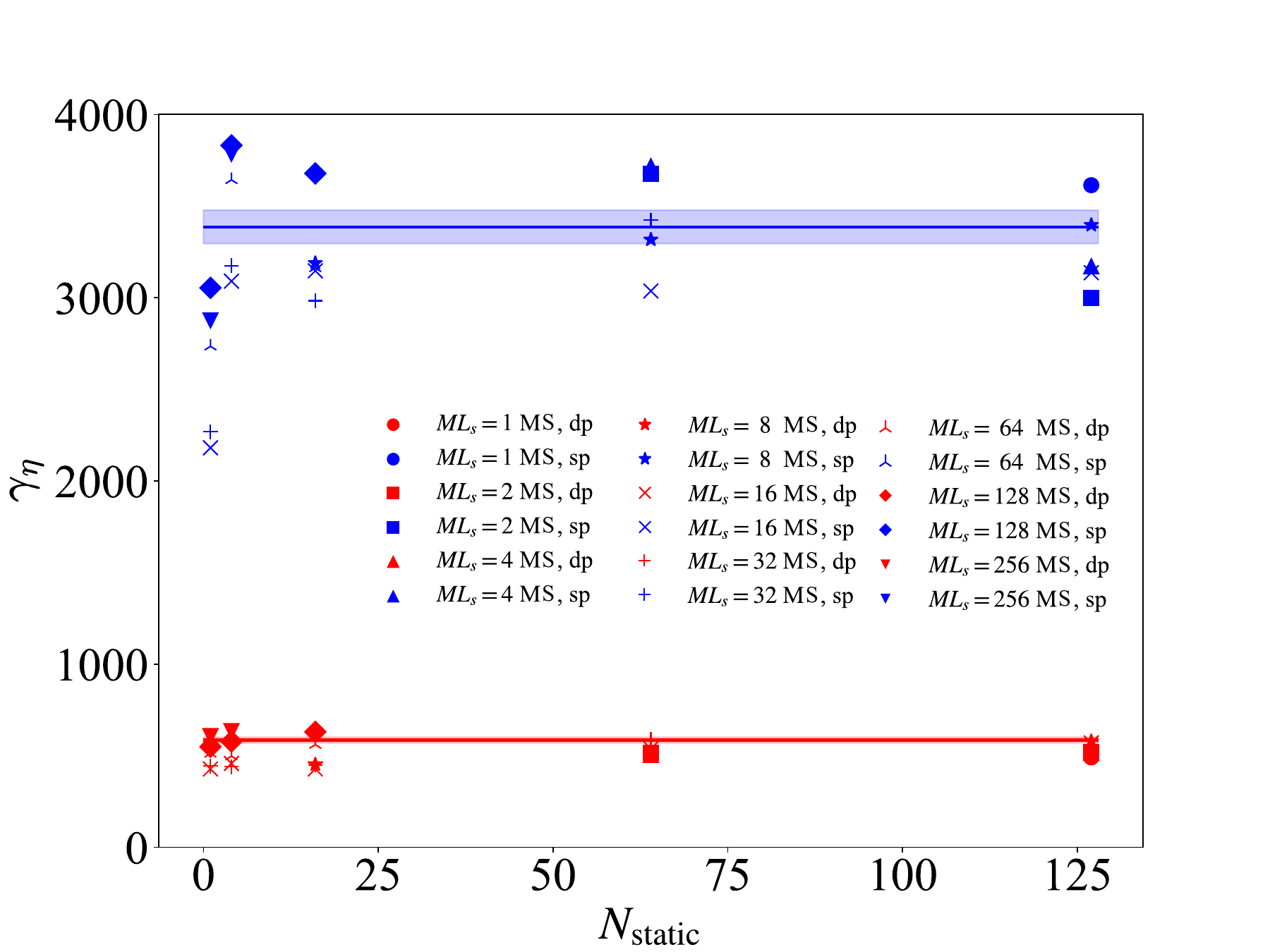}}
    \caption{\label{fig:staticPlot} The speedup ratio $\gamma_\eta$ (where $\eta\equiv\textrm{sp}$ or $\eta\equiv\textrm{dp}$) are plotted as a function of (a) the wavetable length $ML_s$ and (b) the number of static single-frequency tones $N_\textrm{static}$. The lines represent the speedup ratios extracted from fitting the data. Shaded regions represent the errors propagated from the errors in the constant coefficients $C_{\textrm{CPU}}$, $C_{\textrm{GPU,sp}}$ and $C_{\textrm{GPU,dp}}$ extracted from fits. The points represent (measured $t_\text{static,CPU})/{(\textrm{measured }t_{\text{static,GPU},\eta}})$. }
\end{figure}
We verify the static AWG pathway by generating single-tone and multi-tone rf waveforms, and measuring their spectra (see Fig.~\ref{fig:waveforms}). Spectra of example single-tone sinusoidal waveform outputs $V^{\textrm{static}}(t)=a\sin(2\pi f_0 t)$ where $f_0=30$ MHz are presented in Figs.~\ref{fig:chirp2a} and~\ref{fig:chirp2c}. The amplitude $a$ is chosen to be $2.25$ V, which is 90\% of the maximum output level of the DAC \cite{M4i66}. The elements of the wavetables (see Eq.~\ref{eq:dataparallel}) are computed in double-precision floating-point format (dp) or single-precision floating-point format (sp) before being reconstructed in the analog domain using the 16-bit DAC on the M4i.6622-x8 card\footnote{The datatype of the phase argument variable and the amplitude variable of each sine function representing a single-frequency tone are either in both dp or sp. 
}. 

 The spectra of example multi-tone ``defect-free" waveforms are presented in Figs.~\ref{fig:chirp2b} and~\ref{fig:chirp2d}. We call a multi-tone waveform ``defect-free" if it has the following functional form: $V^{\textrm{static}}(t)=a/n\sum^{n-1}_{j=0}\sin(2\pi (f_0+j\Delta f) t+\theta_j)$ where $n$ is the number of single-frequency tones, $a/n$ is the amplitude of each single-frequency tone, $f_0$ is the start frequency, $\Delta f$ is the frequency spacing, and $\theta_j$ are the phases according to Schroeder for crest/peak factor suppression~\cite{1054411}. Essentially, a ``defect-free'' waveform is a multi-tone waveform in which every tone has the same amplitude $(a/n)$ and the frequency spacing between adjacent tones is constant $(\Delta f)$. 
 
 In Figs.~\ref{fig:chirp2b} and~\ref{fig:chirp2d}, we present a 199-tone ``defect-free" waveform output generated for each case where $n=199$, $f_0=1$  MHz, and $\Delta f=500$ kHz. The spectra were measured using Rohde \& Schwarz's FSV signal analyzer (using a resolution bandwidth of 10 Hz and video bandwidth of 3 Hz). Note the deterioration in the spurious free dynamic range (SFDR) in the spectra with wavetable entries calculated in sp over spectra with wavetable entries calculated in dp. Lastly, the wavetables have length $L_s=2^{18}$ (see Sec. \ref{sec:kernels}) for these measurements. 

We also compare the time it takes for the GPU versus the function generator widget on SBench6 6.5.4 build 21020 (the official software provided by the M4i.6622-x8 card vendor Spectrum Instrumentation) running on the CPU to compute the same ``defect-free'' waveforms. To quantify the speedup, we perform the same calculation on Sbench6 and GPU: generate a ``defect-free'' waveform output of the form $V^{\textrm{static}}(t)=b\sum^{N_\textrm{static}}_{j=0}\sin(2\pi(f_0+j\Delta f) t+\theta_j)$ of wavetable length $ML_s$, where $b$ is the amplitude of each single-frequency tone, $M$ is a positive integer, $f_0=50$ MHz, $\Delta f=1$ MHz, and $\theta_j$ are the phases according to Schroeder for crest/peak factor suppression. We vary the wavetable length $ML_s$ and the number of static single-frequency tones $N_{\textrm{static}}$  and record the computation time $t_{\text{static}}$.

The SBench6 function generator computes data in dp. We fit the SBench6 data to the expression $t_{\text{static,CPU}}=C_\text{CPU}N_\text{static}(ML_s)$, where $C_\text{CPU}$ is a constant coefficient. The GPU can compute the wavetable data in sp and dp. We fit the GPU data to the expressions  $t_{\text{static,GPU},\eta}=C_{\text{GPU},\eta}N_\text{static}(ML_s)$, where $ C_{\text{GPU},\eta}$ is a constant coefficient with $\eta\equiv\textrm{sp}$ or $\eta\equiv\textrm{dp}$ representing data computed in sp or dp respectively. Given these expressions, we define the speedup ratio ($\gamma$) as
\begin{equation}
    \gamma_\eta=\frac{t_\text{static,CPU}}{t_{\text{static,GPU},\eta}}=\frac{C_\text{CPU}}{C_{\text{GPU},\eta}}.
\end{equation}

From fitting the data, we get {$C_\textrm{GPU,sp}=0.0252(2)$ ms/MS, $C_\textrm{GPU,dp}=0.146(2)$ ms/MS, $C_\textrm{CPU}=85(1)$ ms/MS.}
The speedup ratio extracted from these fitted coefficients $C_\textrm{GPU,sp}$, $C_\textrm{GPU,dp}$, and $C_\textrm{CPU}$ is $\gamma_{\textrm{sp}}=C_\textrm{CPU}/C_\textrm{GPU,sp}=3389^{+92}_{-90}$ and $\gamma_{\textrm{dp}}=C_\textrm{CPU}/C_\textrm{GPU,dp}=586^{+16}_{-16}$ respectively, which are represented as blue and red lines in Fig.~\ref{fig:staticPlot}. The superscripts and subscripts in the value for $\gamma_{\eta}$ are the asymmetric errors calculated from the coefficient fit errors and are represented by the thickness of the shaded regions in Fig.~\ref{fig:staticPlot}. The points in Fig.~\ref{fig:staticPlot} are the ratio of (measured $t_\text{static,CPU})/{(\textrm{measured }t_{\text{static,GPU},\eta}})$ for different wavetable lengths $ML_s$ and number of static tones $N_{\textrm{static}}$. $\gamma_\eta$ is largely independent of these variables and clearly shows that the static AWG pathway computation on the GPU is a few orders of magnitude faster than the SBench6 computation on the CPU. Lastly, calculating wavetables in sp is faster than calculating in dp at the expense of SFDR and phase noise.

In the next section, we quantify the performance of the dynamic pathways. SBench6 cannot perform on-the-fly computation of dynamic waveforms.

\subsection{Dynamic AWG pathways}

\begin{figure}
    \centering
    \subfloat[\label{fig:chirp3a}]{\includegraphics[width=0.50\linewidth]{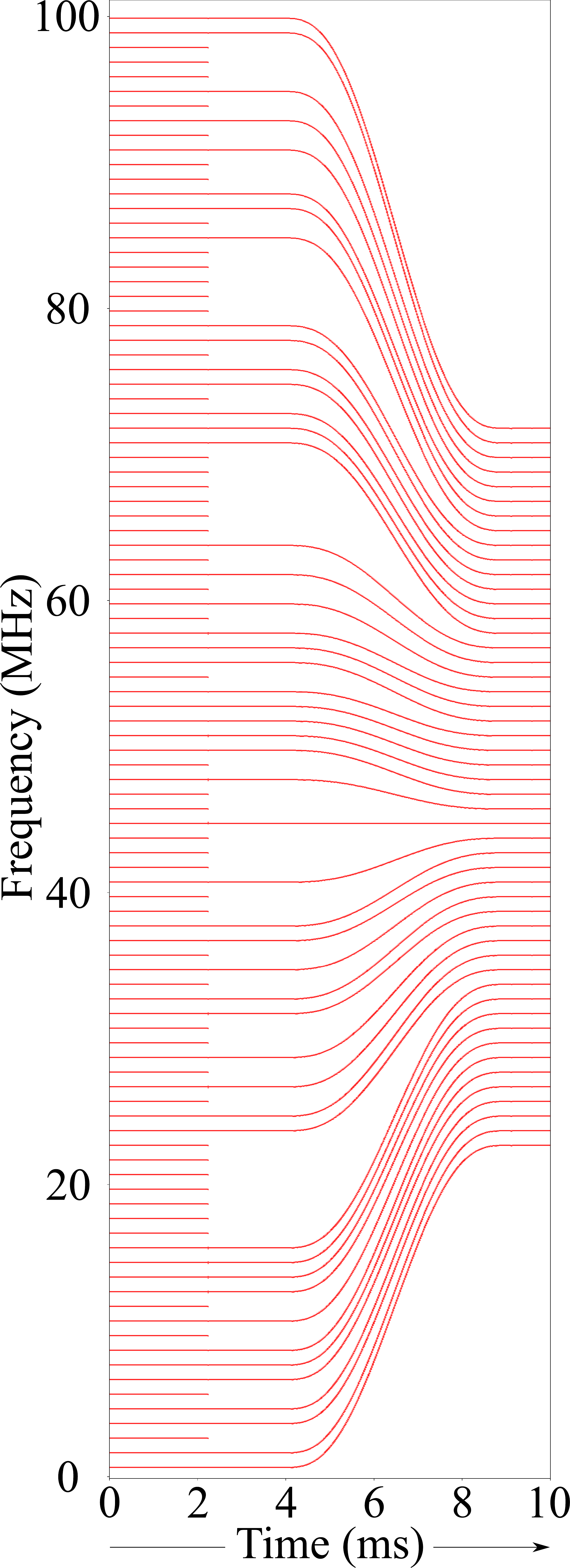}}
    \subfloat[\label{fig:chirp3b}]{\includegraphics[width=0.50\linewidth]{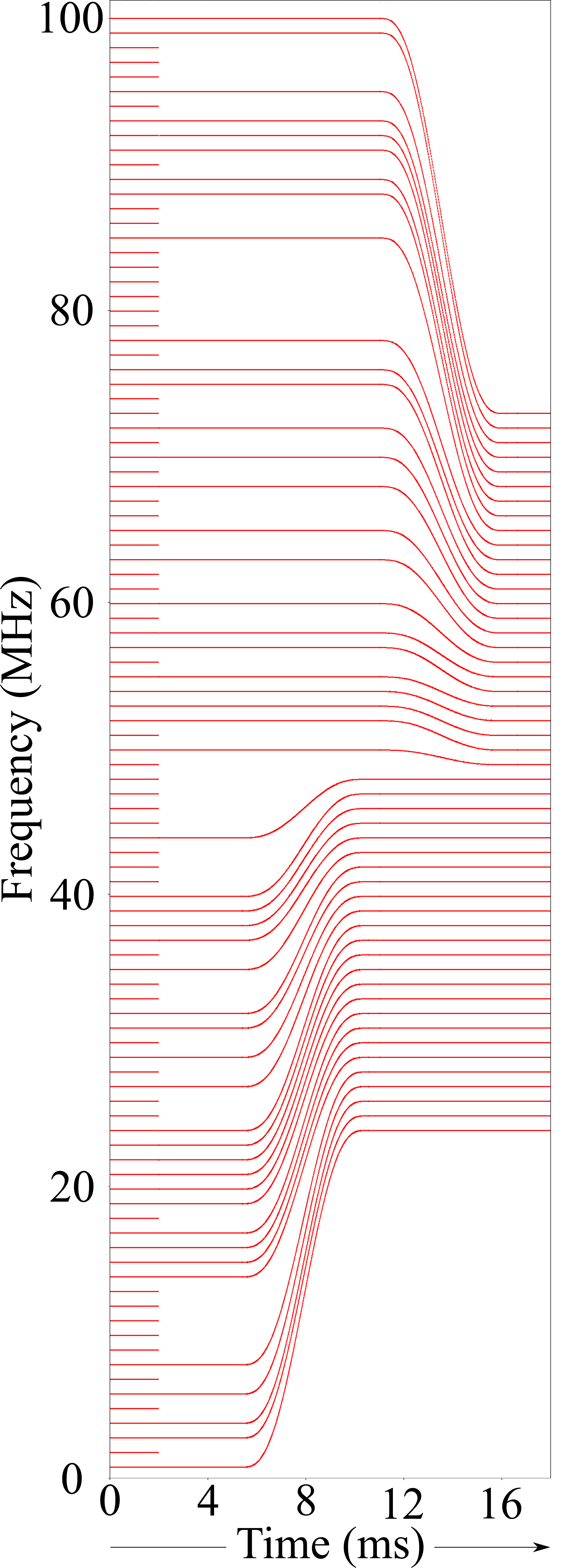}}
    \caption{Measured rf spectrograms simulating the rearrangement of a ``stochastically loaded" array into a ``defect-free" array in the rf domain (slowed-down in time) using the (a) Playback pathway; (b) Streaming pathway. No atoms, laser light, or AODs were used for these demonstrations. We only measure the rf spectrograms in these demonstrations. 
    }
    \label{fig:rearrangement0}
\end{figure}

We verify the dynamic pathways by demonstrating the rearrangement of a ``stochastically loaded'' tweezer array into a ``defect-free'' array in the rf domain. We do not use atoms, laser light, or AODs for this demonstration. We demonstrate the rearrangement using multi-tone rf waveforms, where each single-frequency tone in frequency space maps to an optical tweezer in physical space. 

We demonstrate the rearrangement process using both Playback and Streaming pathways and verify the process by measuring the spectrograms (see Fig.~\ref{fig:rearrangement0}). In Fig.~\ref{fig:chirp3a} and Fig.~\ref{fig:chirp3b}, we present the spectrogram for a typical rearrangement process performed using the Playback pathway and the Streaming pathway respectively. For the demonstrations, we start with a 100-tone ``defect-free" waveform. Then 50\% of the tones are turned off at random, emulating typical atom-loading statistics in optical tweezer array experiments. The remaining tones in this multi-tone waveform with defects representing a stochastically loaded array in the rf domain are then chirped via the minimum jerk trajectory (rearranged) to create a 50-tone ``defect-free" waveform (see Eq.~\ref{eq:freqchirp}). 

In order to measure a spectrogram, we acquire the time trace of the entire waveform using Tektronix's DSA72004B digital serial analyzer (running at a sampling rate of 625 MS/s) for each pathway. Fast Fourier transform (FFT) was performed on contiguous 40 $\mu$s chunks of the acquired waveform time trace to compute the spectrum locally in time. We then stitched the computed spectra together to generate the spectrogram presented in Fig.~\ref{fig:rearrangement0}. The chirping time for each pathway was slowed down to resolve features in the Fourier domain. 

The essential difference between the Playback pathway and the Streaming pathway is that the remaining tones in the multi-tone waveform with defects are chirped in groups in the Streaming pathway while all the tones are chirped simultaneously in the Playback pathway. In Fig.~\ref{fig:chirp3a}, where we demonstrate rearrangement using the Playback pathway, the chirping trajectories for all the remaining tones are calculated on the fly prior to chirping them. In Fig.~\ref{fig:chirp3b}, where we demonstrate rearrangement using the Streaming pathway, the remaining tones are chirped in groups of 25 tones. Once a group of tones finishes chirping, the next group of tones is chirped. This process goes on until all the tones are chirped. While a total of 50 tones were chirped in two groups in this demonstration, the above strategy can be readily extended to chirp a much larger number of tones. We investigate the maximum number of tones that can be chirped in a group in Sec.~\ref{sec:streamingpathway}.

In the following sections, we profile the performance of the dynamic pathways. The initial and final ``defect-free" waveforms are computed in dp unless stated otherwise. We use lower-precision data types in tandem with intrinsic math functions (see the appendix for details) to accelerate the calculation of the dynamic waveform (i.e. the ``rearrangement") that connects the initial and final ``defect-free" waveforms\footnote{
We measured the single-tone and multi-tone spectra using the Rohde \& Schwarz’s FSV signal analyzer and found that the single-tone SFDR and phase noise, and multi-tone SFDR for waveforms calculated using lower-precision data types and intrinsic math functions are better than for waveforms with wavetable table data calculated in sp (Sec.~\ref{sec:Static arbitrary waveform synthesis}). We also performed FFT on the waveform trace acquired on the Tektronix DSA72004B digital serial analyzer to confirm this finding.}. We use the case of chirping the frequencies of single-frequency tones in a multi-tone waveform to profile the performance of the dynamic pathways.  
\subsubsection{Playback pathway}
\begin{figure}[h!]
    \centering
    \subfloat[\label{fig:3dSlice1}]{\includegraphics[width=0.8\linewidth]{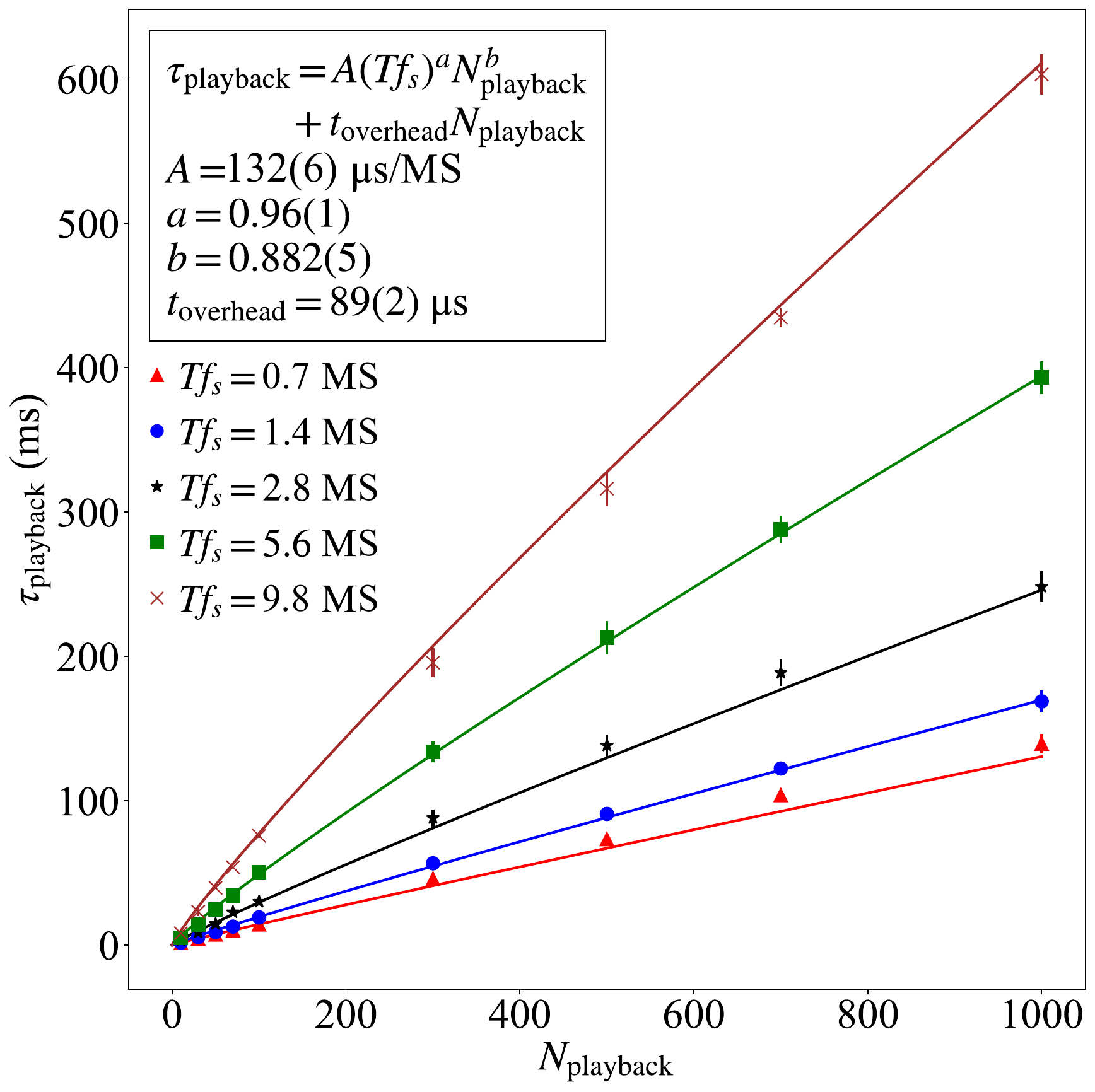}}
    \hfill
    \subfloat[\label{fig:3dSlice2}]{\includegraphics[width=0.8\linewidth]{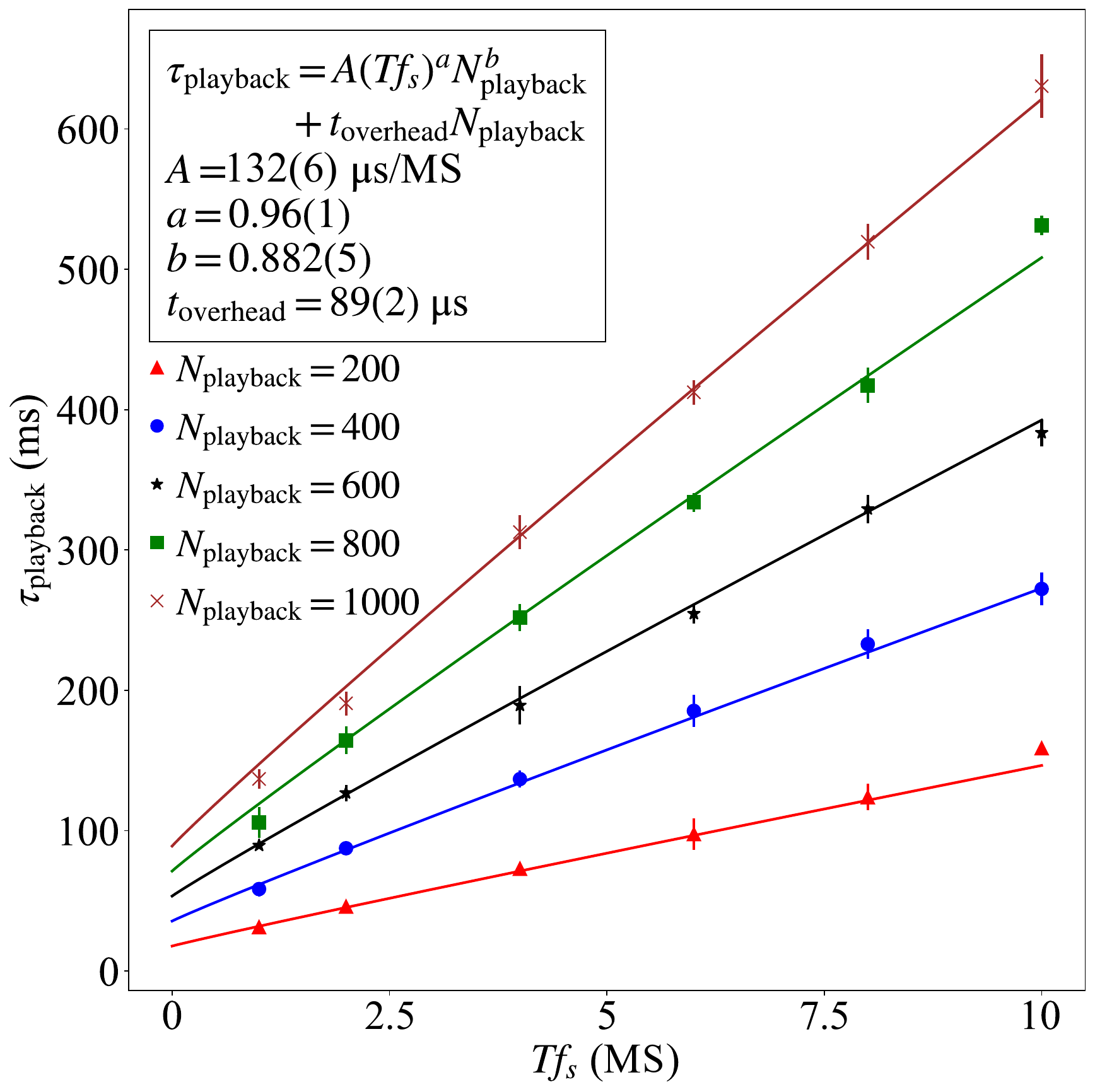}}
    \caption{Modelling the Playback pathway: The measured latency $\tau_{\textrm{Playback}}$  is plotted as a function of (a) the number of chirped tones $N_{\textrm{Playback}}$ and (b) the length of the dynamic waveform $Tf_s$. Points are data and curves are fit to the data.}
    \label{fig:playback_latency}
\end{figure}

The latency from the start of the computation to when the DAC is ready to output the computed real-time arbitrary waveform is quantified by the variable $\tau_{\textrm{playback}}$. The latency depends on the number of chirping tones $N_\mathrm{playback}$ and the length of the dynamic waveform $Tf_s$. Furthermore, the vector addition of $N_\mathrm{playback}$ chirping tones scales as $t_\textrm{overhead}N_\mathrm{playback}$. The Playback pathway computational latency can be modelled as:
\begin{gather}\label{eqn:Playback model}
\tau_{\textrm{playback}}=A(Tf_s)^{a} N_{\textrm{playback}}^{b}+t_\textrm{overhead}N_\mathrm{playback}
\end{gather}
where $A$ is a constant pre-factor; $a,b$ are the dimensionless exponents that capture the scaling dependence on $Tf_s$ and $N_{\textrm{playback}}$ respectively.

We set $f_s= 280$ MS/s for our experiments investigating the latency in the Playback pathway. The latency is extracted using a CPU timer synchronized with the GPU. We fit the data collected from varying $N_{\textrm{playback}}$ and $T$ independently to the two-dimensional model described in Eq.~\ref{eqn:Playback model} (see Fig.~\ref{fig:playback_latency}). In Fig.~\ref{fig:3dSlice1}, we plot the data (represented as points) alongside slices of the global fit along the $N_{\textrm{playback}}$ axis. In Fig.~\ref{fig:3dSlice2}, we plot the data alongside slices of the global fit along the $Tf_s$ axis. Both the exponents $a$ and $b$ are sub-linear. This suggests that with a greater number of $N_{\textrm{playback}}$ tones and larger dynamic waveform lengths $Tf_s$, more blocks are assigned to each SM on the GPU, which helps with the overall GPU throughput via latency hiding. 

Apart from the latency, another limitation inherent to the Playback pathway is the upper bound on the total memory required by the pathway. This memory requirement must not exceed the GPU DRAM capacity. Given the Quadro RTX 6000 GPU's memory capacity of 24 GB, this imposes a constraint on the maximum number of tones that can be chirped, which is approximately 1000 tones chirped for 35 ms at a sampling rate of 280 MS/s. 

\subsubsection{Streaming pathway}
\label{sec:streamingpathway}
\begin{figure}[h]
   \centering
   \subfloat[\label{fig:Streaming performance}]{\includegraphics[width=\linewidth]{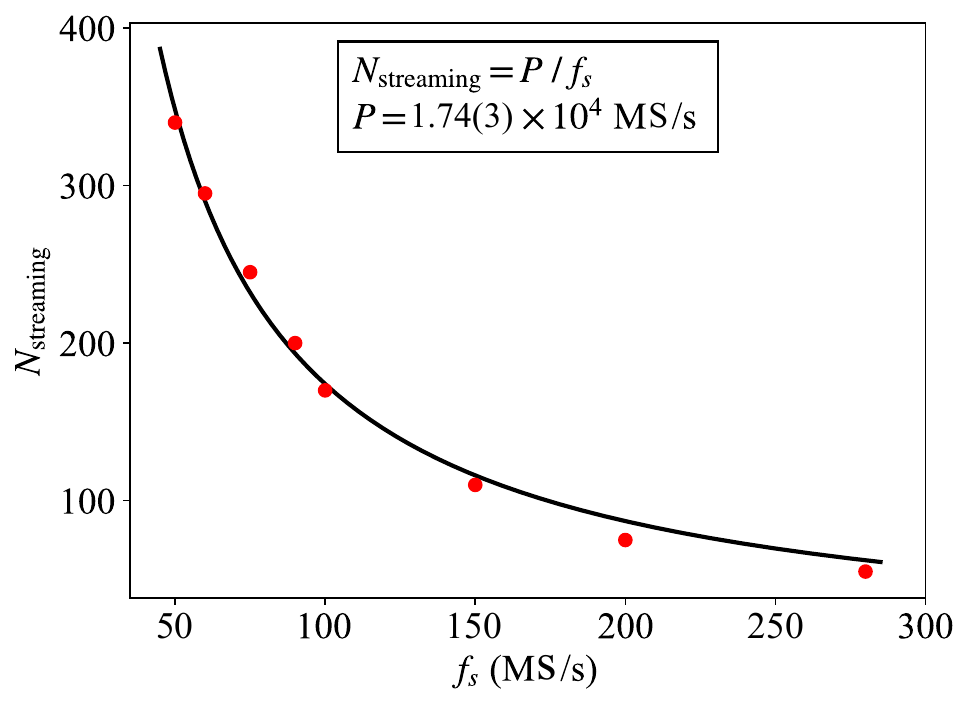}
   }
    \hfill
    \subfloat[\label{fig:latency_streaming}]{\includegraphics[width=\linewidth]{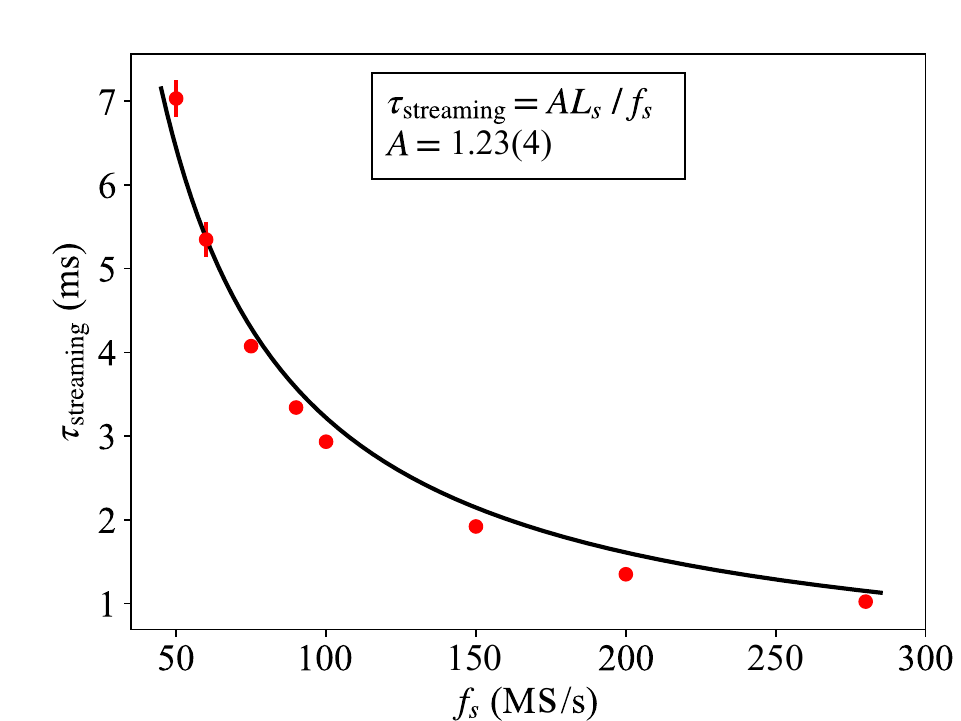}}
    \caption{ Modeling the Streaming pathway: {(a)} Maximum number of tones $N_{\textrm{Streaming}}$ that can be chirped is plotted as a function of the sampling rate $f_s$, (b) the latency $\tau_\textrm{streaming}$ associated with chirping $N_{\textrm{Streaming}}$ tones is plotted as a function $f_s$. Points are data and curves are fit to the data.
    \\
   }
\end{figure}

In the Streaming pathway, each subsequent waveform chunk is computed in the time interval between the current and the next RDMA transfer. This computation time is closely tied to the time it takes to transfer data from the pinned memory buffer of the GPU to the FIFO memory and DAC on the M4i.6622-x8 card.

It takes $L_s/f_s$ amount of time to transfer $L_s$ samples of data in the pinned memory buffer to the FIFO memory and DAC. On the other hand, the time it takes to generate an $L_s-$sized chunk of data representing the dynamic behavior of $N_{\textrm{streaming}}$ chirped tones is 
$N_{\textrm{streaming}}L_s/P$. Here $N_{\textrm{streaming}}L_s$ is the total number of samples that are computed and added together, and $P$ is the rate (in MS/s) at which this computation is performed. Equating both times yields:
\begin{equation}  N_{\textrm{streaming}}=\frac{P}{f_s}.\label{eqn: Streaming model}
\end{equation}
\begin{figure}
    \centering
    \subfloat[\label{fig:chirp1}]{\includegraphics[width=0.88\linewidth]{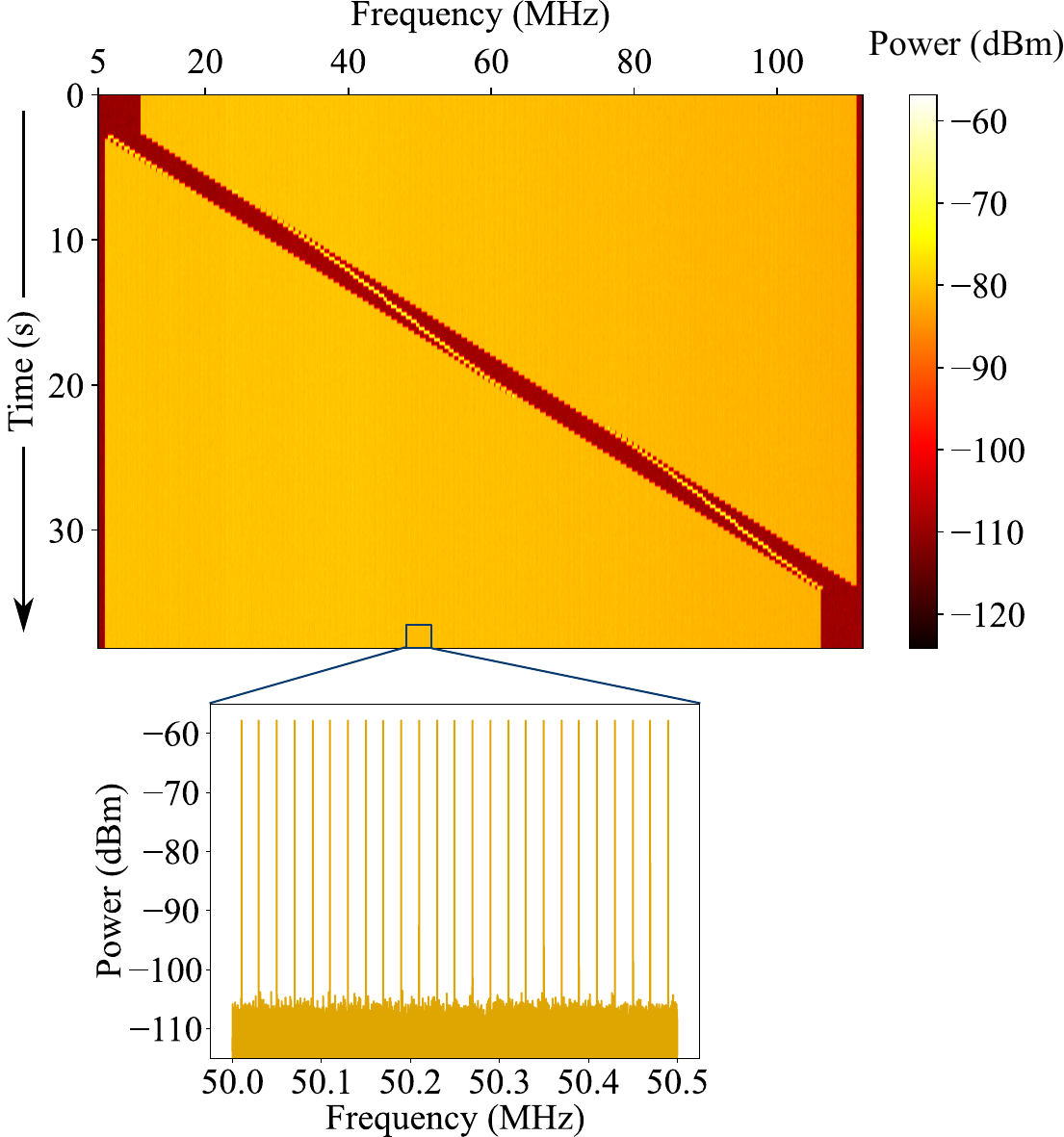}}\hfill
    \subfloat[\label{fig:chirp2}]{\includegraphics[width=0.88\linewidth]{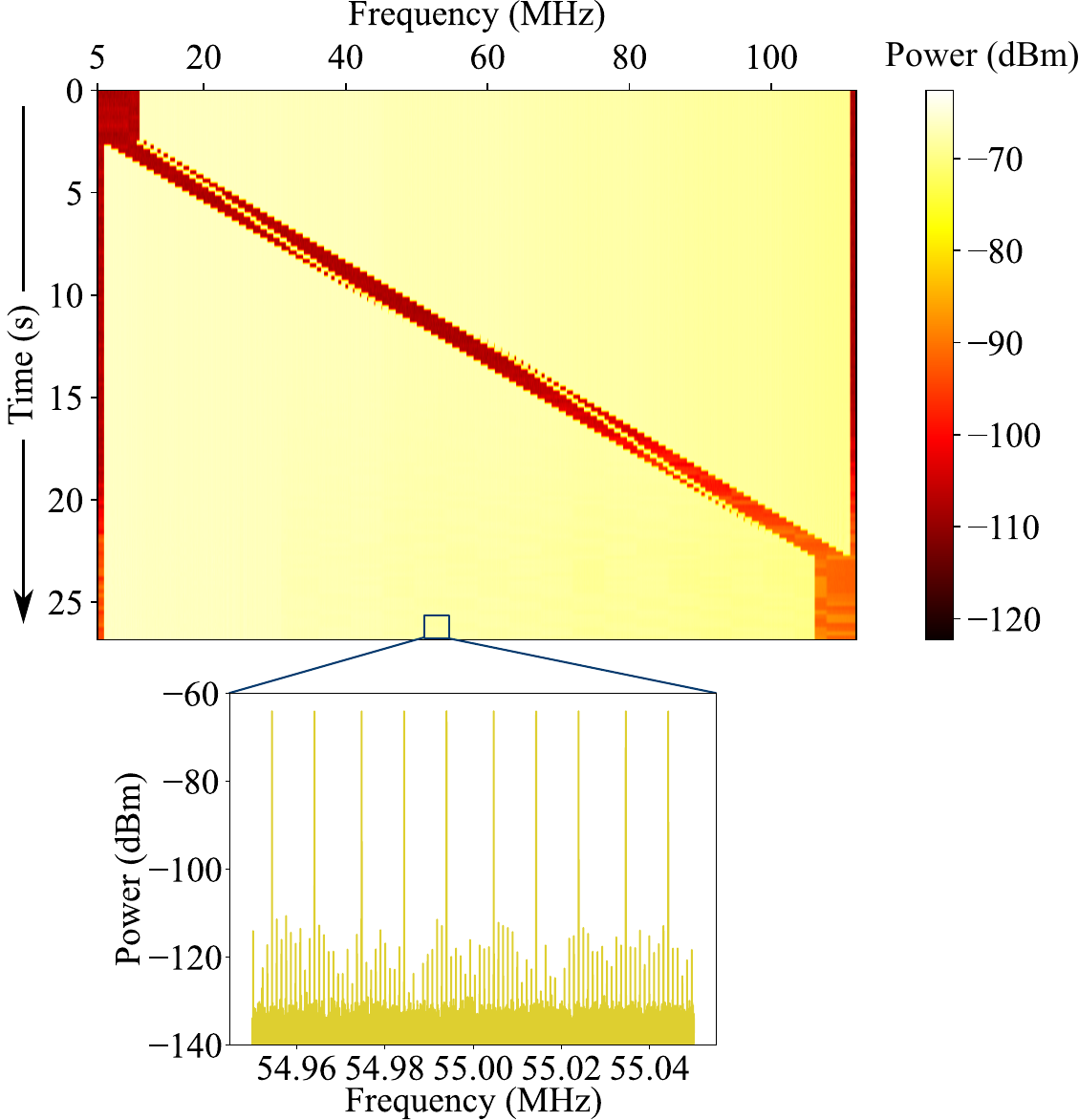}}
    \caption{Chirping large number of tones using the Streaming pathway: (a) Measured spectrogram of a 5000-tone ``defect-free" waveform chirped by $-6$ MHz in groups of 40 tones (initial and final ``defect-free" waveforms are computed in dp) (b) Measured spectrogram of a 10,000-tone ``defect-free" waveform chirped by $-6$ MHz in groups of 100 tones (initial and final ``defect-free" waveforms computed in sp).\\
    }
    \label{fig:rearrangement}
\end{figure}

Given a sampling rate $f_s$, we experimentally determine the maximum number of tones $N_{\textrm{streaming}}$ that can be simultaneously chirped for an indefinite period of time. To determine $N_{\textrm{streaming}}$, we chirp a total of 5000 tones in groups of varying numbers of tones for a particular value of $f_s$. The maximum number of tones that can be chirped together in a single group without the program throwing a buffer underrun error is $N_{\textrm{streaming}}$ for that $f_s$. We then fit these data to the model in Eq.~\ref{eqn: Streaming model} to extract $P$ (Fig.~\ref{fig:Streaming performance}). The quality of the fit implies that the model describes the data well. For initial and final ``defect-free" waveforms computed in dp, we can simultaneously chirp $55$ tones at $f_s=280$ MS/s, and $340$ tones at $f_s=50$ MS/s. For initial and final ``defect-free" waveforms computed in sp, we can chirp $76$ tones at $f_s=280$ MS/s, and $420$ tones at $f_s=50$ MS/s. 

We also measure the latency ($\tau_\textrm{streaming}$) inherent to this pathway. Before the chirping can be initiated, part of the chirping trajectory for the first $N_{\textrm{streaming}}$ tones needs to be computed. This computation is the source of the latency. We model this latency as follows:
\begin{equation}
  \tau_\textrm{streaming}=A \frac{L_s}{f_s},
    \label{eqn:streaming_latency}
\end{equation}
where $A$ is a dimensionless coefficient. 

For our latency measurements, we chirp a 5000-tone waveform in groups of $N_{\textrm{streaming}}$ tones for a particular value of $f_s$. We record $\tau_\textrm{streaming}$ using a CPU timer synchronized with the GPU (Fig.~\ref{fig:latency_streaming}). We then fit this data to Eq.~\ref{eqn:streaming_latency}. Mathematically, $A$ needs to satisfy the inequality $1\le A<1.62$ to prevent buffer overrun and underrun for our FIFO buffer of length $2L_s$\footnote{As latency increases with the size of the FIFO buffer~\cite{anUsingDDS}, we chose to use the smallest buffer size of $2L_s$.}. $A=1.23(4)$, extracted from the fit, satisfies the inequality.  

Last but not least, we use the Streaming pathway to chirp a large number of tones. For initial and final ``defect-free" waveforms evaluated in dp, we chirp a 5000-tone waveform by $-6$ MHz in groups of 40 tones in 250 ms (Fig.~\ref{fig:chirp1} shows the measured spectrogram) at $f_s=280$ MS/s. For initial and final ``defect-free" waveforms evaluated in sp, we chirp a 10,000-tone waveform by $-6$ MHz in groups of 100 tones in 250 ms (Fig.~\ref{fig:chirp2} shows the measured spectrogram) at $f_s=280$ MS/s. The start frequency $f_0$ is 11 MHz and the stop frequency $f_0+n\Delta f$ is 111 MHz for the initial ``defect-free" waveform. The chirping rate was reduced for rf spectrogram acquisition on the FSV signal analyzer (with the resolution bandwidth and video bandwidth set to 3 kHz and 1 kHz respectively). We show the zoomed-in spectrum of the final ``defect-free" multi-tone waveforms for both cases. The two diagonal thin lines between the orange (Fig.~\ref{fig:chirp1}) and yellow (Fig.~\ref{fig:chirp2}) islands arise from aliasing during spectrogram acquisition measurements. The maximum number of tones that can be chirped in a multi-tone waveform is limited by the data size of the initial and final ``defect-free'' waveforms in addition to the GPU specifications, with higher-end GPUs likely allowing for control over a larger number of tones.

\subsection{Other performance metrics}
\subsubsection{Intrinsic latency}
The total latency in the dynamic pathways is the computational latency associated with each pathway ($\tau_{\textrm{playback}}$, $\tau_{\textrm{streaming}}$) plus the latency intrinsic to the GPU-based AWG architecture. The intrinsic latency is the time it takes for the data in the pinned memory of the GPU to be reconstructed in the analog domain by the M4i.6622-x8 card. The worst-case intrinsic latency is 2.6 ms~\cite{emails,anAWGLatency}. 
\subsubsection{Power consumption}
We quantify the GPU power consumption using NVIDIA's system management interface for their GPUs \cite{nvidia-smi}.
The GPU consumes about 74 W of power when performing RDMA transfers or when idle. The power consumption rises to 165 W on average (the maximum recorded value is 173 W) during waveform computation. {In order to perform the measurement during waveform computation, we probed the GPU while it computed the chirping trajectory for a 1000-tone waveform with a chirp duration of 35 ms using the Playback pathway, which takes 610 ms to compute (see Fig.~\ref{fig:3dSlice1}). We programmed the system management interface to probe the power consumption during the computation every 50 ms and ensured that the computation consumed 100\% of the computational resources on the GPU.} According to the datasheet, the M4i.6622-x8 card consumes a maximum of 40 W \cite{M4i66,emails}. Therefore, the worst-case power consumption in the AWG architecture presented here is 205 W. 

For comparison, when the M4i.66xx-x8 or M4i.96xx-x8 card from Spectrum Instrumentation is run in multi-tone DDS mode, it would likely consume a maximum power of 40 W, as mentioned in the device datasheet \cite{M4i66,M4i96}. The CPU is essential in implementing either the commercial multi-tone DDS paradigm using the Spectrum Instrumentation M4i.66xx-x8/M4i.96xx-x8 card or the AWG framework presented here. Therefore, we do not include CPU power consumption in these estimates.

\section{Summary and Outlook}

In this paper, we have implemented a novel fast real-time arbitrary waveform generation architecture using additive synthesis. Casting additive synthesis as an ``embarrassingly'' data-parallel problem allowed us to solve it efficiently on a GPU. We developed the software using the CUDA API from NVIDIA. We used a Quadro RTX 6000 GPU and an M4i.6622-x8 card from Spectrum Instrumentation mounted on the motherboard of a personal computer as the hardware for a proof-of-principle demonstration of the architecture. 

We implemented multiple AWG pathways: one static pathway, and two dynamic pathways. We used the static AWG pathway to generate periodic arbitrary waveforms. We demonstrated a 586-fold speedup in the computation time of these waveforms compared to the CPU. The dynamic AWG pathways are classified according to their ability to (a) generate real-time arbitrary waveforms for a short duration of time (Playback pathway), or (b) indefinitely stream less complex real-time arbitrary waveforms (Streaming pathway). We demonstrated the
rearrangement of a multi-tone waveform with defects
into a ``defect-free'' multi-tone waveform using both pathways. We measured and characterized the latency inherent to both dynamic pathways. In the Streaming pathway, we measured the latency to be in the order of a few milliseconds. Also using the Streaming pathway, we demonstrated control of over 5000 tones by chirping the tones in groups.

Our AWG architecture has a few limitations. When we profiled our implementations on NVIDIA Nsight, it stated that data transfer between DRAM memory and caches limited the performance of the programs. Therefore, memory access patterns inside the GPU may be further optimized. We believe that using better GPUs may alleviate this problem. For instance, NVIDIA's RTX A6000 \cite{RTXA6000}, which is compatible with our AWG architecture~\cite{emails}, has a greater memory access bandwidth (768 GB/s) than the RTX 6000 (up to 672 GB/s) used here. RTX A6000 also has double the DRAM capacity of Quadro RTX 6000 and better single-precision performance (38.7 TFLOPS versus 16.3 TFLOPs).

The RDMA transfer speed in our AWG architecture is currently limited by the PCIe 2.0 implementation on the Virtex 6 FPGA on the M4i.6622-x8 card. The RDMA transfer speeds can be improved by implementing faster PCIe protocols. This AWG architecture is also limited by the GPU's processing speed, memory, and high-speed access to the memory by streaming multiprocessors. Given that the speed of floating-point operations and the number of cores have been increasing steadily with every GPU generation, we expect our AWG architecture to be further accelerated in the future. Additionally, multiple GPUs may be connected together via NVLink and NVSwitch to further accelerate additive synthesis. CUDA libraries such as cuBLAS and cuFFT can be readily integrated into our software for more complex signal-processing tasks.

We implemented this AWG architecture using commercial off-the-shelf components without any FPGA gateware development. Developing hardware that blends the benefits of CPU, FPGA, and GPU processing architectures~\cite{fpgaforsoftprog} will likely have useful applications. One promising direction is to interface AMD UltraScale+ RFSoC chips \cite{AMDRFSoC} with GPU cards. Lastly, while this architecture was designed with neutral atom quantum computation and simulation in mind, it may be used for radar and built-in self-test~\cite{8103508}. 

{\appendix
\setcounter{secnumdepth}{0}
\section{Appendix: Parallel computation optimization}
\label{sec:Appendix}
We primarily consulted NVIDIA’s CUDA C++ Programming Guide \cite{CUDAGuide} and CUDA C++ Best Practices Guide~\cite{CUDABestPractice} to boost computational efficiency. In our implementations, we aimed to minimize computational latency during real-time waveform generation. In fact, we tried to move all the long latency operations to the initialization state.
\begin{itemize}
\item To circumvent bulky data transfers between the GPU and CPU, we allocate all data buffers in GPU memory and reuse temporary data, thus alleviating constraints imposed by PCIe bus bandwidth. All synthesis parameters are transmitted from the CPU to the GPU only during initialization or when static arbitrary waveforms are being reconstructed. This is because the GPU is idle in these states. Besides sending parameters, the CPU exclusively interacts with the GPU when launching CUDA kernel functions.
\item SFUs on the GPU are capable of very fast execution of special floating-point arithmetic functions called intrinsic functions. However, the SFU can only evaluate intrinsic math functions at sp or lower. Additionally, current GPUs are optimized to process vector structures such as float2 vectors or half2 vectors. We use half2 structures in conjunction with intrinsic functions to speed up the computation. 
\item  We store pre-calculated and reusable data in GPU DRAM, such as wavetables, prior to additive synthesis. This offloads computationally intensive tasks to the initialization stage, mitigating the need for repeating computations, which helps reduce latency.
\item Computing the sine of a large number can incur significant computational cost, especially when processing sample data with a large index $i$. To alleviate this problem, we apply a \verb|modf()| to the phase $\phi[i]$ before executing any trigonometric operations:
\begin{align}
    V[i]=\sum_j a_j\sin\left[2\pi\times\textbf{modf}\left(\frac{\phi_j[i]}{2\pi}\right)\right]
\end{align}
\item We use fused multiply–add (FMA) operations and \twound restrict\twound ~qualifier to further improve code performance. The \twound restrict\twound ~keyword declares an array argument of a kernel function to be unaliased, which means the array is only accessible through a single symbol in the scope of the executing kernel. This allows for additional optimization during compilation. Given that our kernels access data in the buffers only once, most arrays in our program benefit from this optimization.

\end{itemize}
}

\section*{Acknowledgment}
The authors thank Alessandro Restelli, Sayantan Sarkar, Yanda Geng, and Alexander Craddock for carefully reading the manuscript. We also thank Alessandro Restelli for letting us borrow the Rohde \& Schwarz FSV signal analyzer and Tektronix DSA72004B digital serial
analyzer. 

{
\bibliographystyle{IEEEtran}
\bibliography{references}
}

\vfill

\end{document}